\documentclass[11pt]{article}
\usepackage{jheppub}
\usepackage{bm,bbm}
\usepackage{booktabs,amsmath}
\usepackage{accents}
\usepackage{multirow}
\usepackage{graphicx}
\usepackage{braket}
\usepackage{mathtools}
\usepackage{comment}
\usepackage{autobreak}
\usepackage{subcaption}
\usepackage{enumerate}
\usepackage{blkarray}
\usepackage{longtable}

\usepackage{adjustbox}

\usepackage[dvipsnames]{xcolor}

\definecolor{defectblue}{RGB}{199,224,231}
\definecolor{bulkgray}{RGB}{245,245,245}
\definecolor{interyellow}{RGB}{246,234,200}

\usepackage{tikz}
\usetikzlibrary{patterns}
\usetikzlibrary{arrows,shapes,positioning}
\usetikzlibrary{decorations.markings}
\usetikzlibrary{decorations.pathmorphing}
\usetikzlibrary{calc}
\tikzset{snake it/.style={decorate, decoration=snake}}

\hypersetup{
    colorlinks=true,
    linkcolor=blue,
    filecolor=magenta,      
    urlcolor=cyan,
    pdfpagemode=FullScreen,
    }

\usepackage{arydshln}
\usepackage{mathtools}

\edef\restoreparindent{\parindent=\the\parindent\relax}
\usepackage{parskip}
\restoreparindent
\usepackage{ascmac}
\usepackage{mathrsfs}

\usepackage{amsthm}
\newtheoremstyle{break}
  {\topsep}{\topsep}%
  {\upshape}{}%
  {\bfseries}{}%
  {\newline}{}%
\theoremstyle{break}

\def\d{{\rm d}}
\def\i{{\rm i}}
\def\CD{{\cal D}}

\def\CF{{\cal F}}

\def\CM{{\cal M}}

\def\CT{{\cal T}}

\def\BC{\mathbb{C}}

\def\BH{\mathbb{H}}
\def\BR{\mathbb{R}}
\def\BS{\mathbb{S}}
\def\BT{\mathbb{T}}
\def\BQ{\mathbb{Q}}
\def\BZ{\mathbb{Z}}

\def\d{\mathrm{d}}
\def\SO{\mathrm{SO}}

\def\U{\mathrm{U}}

\newcommand{\hodge}{{\ast}}

\title{Symmetry fractionalization and duality defects in Maxwell theory}
    
\author[a]{Naoto Kan,}
\author[b]{Kohki Kawabata}
\author[a]{and Hiroki Wada}

\affiliation[a]{Department of Physics, Osaka University,\\
Machikaneyama-Cho 1-1, Toyonaka 560-0043, Japan}

\affiliation[b]{Department of Physics, Faculty of Science,
The University of Tokyo,\\
Bunkyo-Ku, Tokyo 113-0033, Japan}

\preprint{OU-HET-1232}

\abstract{
We consider Maxwell theory on a non-spin manifold. Depending on the choice of statistics for line operators, there are three non-anomalous theories and one anomalous theory with different symmetry fractionalizations.
We establish the gauging maps that connect the non-anomalous theories by coupling them to a discrete gauge theory.
We also construct topological interfaces associated with $\mathrm{SL}(2,\mathbb{Z})$ duality and gauging of electric and magnetic one-form symmetries.
Finally, by stacking the topological interfaces, we compose various kinds of duality defects, which lead to non-invertible symmetries of non-spin Maxwell theories.
}

\begin{document}
\maketitle
\flushbottom

\newpage

\section{Introduction}
Maxwell theory in four dimensions serves as a testing ground for generalized symmetries. 
The recent progress in the generalization of global symmetries reveals two important concepts: higher-form symmetry~\cite{Kapustin:2014gua,Gaiotto:2014kfa} and non-invertible symmetry~\cite{Bhardwaj:2017xup,Tachikawa:2017gyf,Chang:2018iay}.
The higher-form symmetry generalizes ordinary symmetry such that its action is on higher-dimensional extended objects and provides a powerful tool for elucidating the dynamics of strongly coupled theories~\cite{Gaiotto:2017yup}.
On the other hand, the non-invertible symmetry is another generalization of global symmetry with fusion rules beyond the group-theoretical framework and plays a critical role in constraining the renormalization group flow~\cite{Chang:2018iay}.
Despite its simple appearance, the Maxwell theory encapsulates both two generalized symmetries.
The theory has electric and magnetic one-form symmetries, whose charged objects are Wilson and 't~Hooft lines.
By combining the gauging of the one-form symmetries with the electric-magnetic duality, it turns out to possess non-invertible symmetries at specific values of the coupling constant.
Their profiles such as the fusion rule and the action on physical operators are extensively studied in~\cite{Choi:2021kmx,Choi:2022zal,Choi:2022rfe,Niro:2022ctq,Cordova:2023ent} (see also~\cite{Kapustin:2010if,Cordova:2019wpi,Thorngren:2019iar,Ji:2019jhk,Kong:2020cie,Rudelius:2020orz,Gaiotto:2020iye,Komargodski:2020mxz,Nguyen:2021yld,Lin:2021udi,Inamura:2021wuo,Nguyen:2021naa,Heidenreich:2021xpr,Ji:2021esj,Thorngren:2021yso,Delmastro:2021otj,Kong:2021equ,Sharpe:2021srf,Koide:2021zxj,Huang:2021zvu,Kaidi:2021xfk,Roumpedakis:2022aik,Hayashi:2022fkw,Kaidi:2022uux,Choi:2022jqy,Cordova:2022ieu,Bhardwaj:2022lsg,Bartsch:2022mpm,Heckman:2022muc,Kaidi:2022cpf,Chen:2022cyw,Bashmakov:2022uek,Cordova:2022fhg,Choi:2022fgx,Yokokura:2022alv,Bhardwaj:2022kot,Bhardwaj:2022maz,Bartsch:2022ytj,Heckman:2022xgu,Apte:2022xtu,Kaidi:2023maf,Koide:2023rqd,Damia:2023ses,Bhardwaj:2023ayw,vanBeest:2023dbu,Lawrie:2023tdz,Apruzzi:2023uma,Chen:2023czk,Pace:2023kyi,Cordova:2023bja,Antinucci:2023ezl,Benedetti:2023owa,Choi:2023pdp,Nagoya:2023zky,vanBeest:2023mbs,Okada:2024qmk} for related developments).

The characterization of Maxwell theory is significantly influenced by the spacetime manifold.
On a spin manifold, the charge lattice of line operators uniquely defines Maxwell theory.
However, on a non-spin manifold, Maxwell theory has ambiguity even after specifying the charge lattice, which stems from symmetry fractionalization~\cite{Barkeshli:2014cna,Zou:2017ppq,Cordova:2017kue,Cordova:2018acb,Hsin:2019gvb,Wang:2019obe,Yu:2020twi,Delmastro:2022pfo,Ye:2022bkx,Brennan:2023kpo,Brennan:2023tae,Brennan:2023ynm,Brennan:2023vsa,Hsin:2024eyg}.
In non-spin Maxwell theories, the symmetry fractionalization dictates that some line operators have a half-integer spin under the Lorentz group $\mathrm{SO}(4)$ and become fermionic.
To define the theory without ambiguity, we need to specify the symmetry fractionalization class, equivalently the choice of statistics for the set of line operators~\cite{Metlitski:2015yqa,Hsin:2019fhf,Ang:2019txy,Brennan:2022tyl,Davighi:2023luh}.
This yields four options after giving the charge lattice (see Fig.~\ref{fig:statistics_line}).
In other words, we have four consistent non-spin Maxwell theories with different symmetry fractionalizations.
Among them, three are non-anomalous, and the other has a pure gravitational anomaly and is called all-fermion electrodynamics~\cite{Wang:2013zja,Thorngren:2014pza,Kravec:2014aza,Wang:2018qoy}.

In this paper, we establish the interrelationships between non-spin Maxwell theories by gauging one-form symmetries.
To capture their relations, we construct the maps between the theories with different symmetry fractionalizations, called the symmetry fractionalization maps~\cite{Hsin:2019gvb}.
Since a non-trivial fractionalization class is realized by turning on an appropriate background field of one-form symmetries, we modify the background field by gauging a one-form symmetry with a four-dimensional discrete gauge theory called BF theory. To see it more concretely, suppose that $\CT$ and $\CT'$ are two Maxwell theories. Then, we map $\CT$ to $\CT'$ by gauging the diagonal $\BZ_2$ symmetry with the BF theory:
\begin{align}
    \CF:\;\CT \;\longmapsto \;\frac{\CT \times_{\rm e,m} \text{BF}[C_1,C_2]}{\BZ_2}\; \cong\; \CT'\,,
\end{align}
where $\mathrm{e,m}$ denote the coupling by electric and magnetic one-form symmetries, respectively.
Depending on the background gauge fields $C_1$ and $C_2$, we have three choices of BF theory.
After gauging, the background gauge fields in the theory $\CT$ are converted into the ones of $\CT'$, and we obtain the symmetry fractionalization map $\CF: \CT\mapsto \CT'$.
See Fig.~\ref{fig:fm} for the result of symmetry fractionalization maps between non-spin Maxwell theories.
Note that we restrict our attention to the three non-anomalous Maxwell theories to define the $\BZ_2$ gauging consistently.

To reveal the duality structure on a non-spin manifold, we construct topological interfaces in Maxwell theories.
Specifically, we consider $\mathrm{SL}(2,\BZ)$ duality interfaces and gauging interfaces.
The $\mathrm{SL}(2,\BZ)$ duality is generated by the electric-magnetic duality and the periodicity of theta parameter and ensures equivalence between the theories with different coupling constants.
On a non-spin manifold, Maxwell theories covariantly transform under $\mathrm{SL}(2,\BZ)$ duality as in Fig.~\ref{fig:duality}, so the interfaces typically connect the theories with different symmetry fractionalizations.
We explicitly write down the $\mathrm{SL}(2,\BZ)$ interface actions by imposing the equations of motion at the interface and the continuity condition of the energy-momentum tensor for their topological nature.
Furthermore, we construct the gauging interfaces utilizing half gauging construction~\cite{Choi:2021kmx}.
Associated with each symmetry fractionalization map $\CF:\CT\mapsto \CT'$, we show the action of the corresponding interface that glues $\CT$ and $\CT'$, and check its invariance under infinitesimal deformation of the interface locus.

Finally, we compose non-invertible duality defects in non-spin Maxwell theories by stacking the topological interfaces together.
We find composite operations from the $\mathrm{SL}(2,\BZ)$ duality and the gauging of the one-form symmetries, under which a theory is self-dual at a specific value of the coupling constant. 
This ensures the existence of symmetries in the theory because we can construct a topological defect by fusing the topological interfaces associated with the self-dual composite operation.
We identify the defects that exist on a spin manifold but do not exist on a non-spin manifold for each of the three non-spin Maxwell theories.
We also demonstrate the construction of topological defects at some specific coupling constants.
By piercing line operators into the topological defects, we show that some of them act non-invertibly on line operators.
At a certain value of the coupling constant, we see that while Maxwell theory with trivial symmetry fractionalization has a duality defect, the theories with non-trivial symmetry fractionalizations do not contain the duality defect.
This suggests the distinction of symmetries between the theories with different symmetry fractionalizations.

The rest of this paper is organized as follows.
In section~\ref{sec:nonspin}, we review Maxwell theories and BF theory on a non-spin manifold.
We explain that the symmetry fractionalization can be understood in terms of background gauge fields of the one-form symmetries.
For later convenience, we carefully discuss the coupling between Maxwell theories and BF theory at the end of this section.
In section~\ref{sec:frac_map}, we explore the symmetry fractionalization maps on a non-spin manifold.
At the beginning of the section, we summarise the results and proceed to their detailed constructions.
Based on the fractionalization maps and $\mathrm{SL}(2,\BZ)$ duality, we construct non-invertible defects in section~\ref{sec:top_defect}.
To this end, we first prepare the topological interfaces associated with the duality and proceed to the gauging of one-form symmetries.
We conclude in section~\ref{sec:discussion} with some discussions.

\section{Coupling to background fields}
\label{sec:nonspin}

In this section, we consider coupling to background gauge fields on a non-spin manifold, with a particular focus on Maxwell theory and a topological discrete gauge theory. 
In section~\ref{ss:Maxwell}, we review Maxwell theory on a non-spin manifold and show the existence of freedom in the choice of the statistical pattern of line operators.
Section~\ref{ss:BF_theory} describes a topological $\BZ_n$ gauge theory, which is extensively used in the later section.
We also discuss the coupling between Maxwell theory and the $\BZ_n$ gauge theory.
After the coupling, the theory has one-form symmetries, and we activate the corresponding two-form background gauge fields using a Stiefel--Whitney class of the tangent bundle.

\subsection{Maxwell theory on non-spin manifold}
\label{ss:Maxwell}

This subsection reviews Maxwell theory on a non-spin manifold following~\cite{Ang:2019txy,Brennan:2022tyl}.
Maxwell theory has a set of line objects associated with the electric and magnetic one-form symmetries, such as Wilson, 't~Hooft, and dyonic operators.
On a non-spin manifold, there is freedom in the choice of the statistics of the line operators, which results in four consistent Maxwell theories with the same coupling constant.
We explain that the four Maxwell theories differ in their choice of background gauge fields for the electric and magnetic one-form symmetries.
Finally, we describe $\mathrm{SL}(2,\BZ)$ duality transformations between the non-spin Maxwell theories.

\subsubsection{Electric and magnetic one-form symmetry}

We define Maxwell theory without charged matters by the Euclidean action
\begin{align}
\label{eq:action_maxwell}
    S = \frac{1}{2e^2}\int_\CM F\wedge * F + \frac{\i\,\theta}{8\pi^2} \int_\CM F\wedge F \,,
\end{align}
where $F = \d A$ is the field strength, $e$ is the coupling constant, and $\theta$ is the theta angle parameter. The first term is a standard kinetic term, and the second is the topological theta term.
We will denote the complex coupling constant by
\begin{align}
    \tau = \frac{\theta}{2\pi} + \frac{2\pi\i}{e^2}\,,
\end{align}
which takes values in the upper half plane $\BH=\{z\in\BC\mid \mathrm{Im}(\tau)>0\}$.
The equation of motion sets $\d \hodge F = 0$, and the Bianchi identity implies $\d F = 0$.

The theory possesses the Euclidean spacetime symmetry $\mathrm{SO}(4)$ and does not require the spacetime manifold $\CM$ to admit a spin structure.
In other words, we can define the theory on a spacetime manifold such that the second Stiefel--Whitney class of the tangent bundle does not vanish: $w_2(\CM)\neq0$.
Throughout this paper, we assume that the spacetime manifold is orientable: the first Stiefel--Whitney class $w_1(\CM)$ is trivial.
Additionally, we suppose the spacetime manifold is closed for a while.
Note that the periodicity of Maxwell theory is $\theta\sim \theta + 2\pi$ on a spin manifold, while the periodicity becomes $\theta\sim\theta + 4\pi$ in the case of a non-spin manifold.\footnote{The periodicity of the theta parameter depends on the type of non-spin Maxwell theories. While the theory defined by \eqref{eq:action_maxwell} has the periodicity $\theta\sim \theta+4\pi$, other theories called $W_\mathrm{f}T_\mathrm{b}$ and $W_\mathrm{f}T_\mathrm{f}$ in a later section have $\theta\sim \theta+2\pi$.}

Although Maxwell theory does not have any intrinsic zero-form continuous global symmetry,\footnote{Maxwell theory has the discrete charge-conjugation symmetry and for certain theta-angles, it also has parity and time-reversal symmetries.} it has electric and magnetic one-form global symmetries $\mathrm{U}(1)_e\times\mathrm{U}
(1)_m$~\cite{Gaiotto:2014kfa}.
The electric one-form symmetry acts as
\begin{align}
    A \to A + \lambda_e\,,
\end{align}
where $\lambda_e$ is a flat connection and its current is written as $j_e = (2/e^2)\, \hodge F$, which conserves due to the equation of motion. 
The magnetic one-form symmetry acts on the dual gauge field in the same manner and the conserved current is $j_m = F/2\pi$ guaranteed by the Bianchi identity.
The associated charged operators are Wilson lines and 't~Hooft lines.
A Wilson line with charge $n$ along a curve $\gamma$ is 
\begin{align}
    W^{n} = \exp\left(\i \,n \int_\gamma A\right)\,.
\end{align}
This reflects the worldline of an infinitely heavy particle with electric charge $n$.
We conventionally denote the fundamental Wilson line with $n=1$ as $W$.
Similarly, an 't~Hooft line $T^m$ is the worldline of an infinitely heavy monopole with magnetic charge $m$, and the fundamental 't~Hooft line is $T$ for short.
A general dyonic operator is the product of the fundamental Wilson lines and the fundamental 't~Hooft lines: $W^nT^m$.

We can introduce the two-form background gauge fields $B_e$ and $B_m$ associated with the electric and magnetic one-form symmetry $\U(1)_e\times\U(1)_m$, respectively, by the action
\begin{align}
\begin{aligned}
\label{eq:maxwell_background}
    S=\frac{1}{2e^2} \int_\CM \left|F-B_e\right|^{2}+ \frac{\i\,\theta}{8\pi^2}\int_\CM (F-B_e)^{2} + \frac{\i}{2\pi} \int_\CM (F-B_e)\wedge B_m \,.
\end{aligned}
\end{align}
Here, we use the notation $|X|^2 := X\wedge\hodge X$ and $X^2:=X\wedge X$.
The gauge transformation is given by
\begin{align}
\label{eq:gauge_transf}
    A\to A + \lambda_e\,,\qquad B_e\to B_e + \d\lambda_e\,,\qquad B_m \to B_m + \d \lambda_m\,,
\end{align}
where $\lambda_{e,m}$ is an ordinary $\U(1)$ connection with a period $\int_\Sigma \d\lambda_{e,m}\in2\pi\BZ$ for a two-cycle $\Sigma$ on $\CM$.
Note that the action~\eqref{eq:maxwell_background} is not invariant under the gauge transformation, which implies the existence of mixed 't~Hooft anomaly between electric and magnetic one-form symmetries.
Also, we can gauge only a finite abelian subgroup of the electric and magnetic one-form symmetries
\begin{align}
    \BZ_k^e \times \BZ_{k'}^m \subset \U(1)_e\times\U(1)_m\,.
\end{align}
In this case, the background gauge fields become flat: $\d B_e = \d B_m =0$.
For ${\rm gcd}(k,k')\neq 1$, we again encounter the mixed 't~Hooft anomaly.

\subsubsection{Lorentz symmetry fractionalization}
\label{sss:lorentz}

On a non-spin manifold, Maxwell theory has freedom in the choice of the statistics of line operators. 
This is a notable example of symmetry fractionalization: the phenomenon in which different quantum numbers arise between local operators and line operators under an ordinary global symmetry (see~\cite{Barkeshli:2014cna,Zou:2017ppq,Cordova:2017kue,Cordova:2018acb,Hsin:2019gvb,Yu:2020twi,Delmastro:2022pfo,Ye:2022bkx,Brennan:2023kpo,Brennan:2023tae,Brennan:2023ynm,Brennan:2023vsa,Hsin:2024eyg} for recent progress).
More concretely, while local operators are in a genuine representation of a global symmetry, line operators are in a projective representation.
In the case of the Maxwell theories on a non-spin manifold, the spacetime symmetry group $\SO(4)$ is fractionalized and line operators are in a projective representation of $\SO(4)$, equivalently, a genuine representation of $\mathrm{Spin}(4)$.

Let us focus on the symmetry fractionalization by the spacetime symmetry group $\SO(4)$ (Lorentz symmetry fractionalization~\cite{Hsin:2019gvb}).
When caring about symmetry fractionalization, the global structure of $\SO(4)$ takes importance.
Consider the central extension of $\SO(4)$ following the short exact sequence:
\begin{align}
    1\longrightarrow \BZ_2 \longrightarrow \mathrm{Spin}(4) \longrightarrow \SO(4) \longrightarrow 1\,,
\end{align}
where $\mathrm{Spin}(4)$ is the universal cover of $\SO(4)$ and $\BZ_2$ is the fermion parity.
The projective representation is classified by the second group cohomology $H^{2}(\SO(4),\BZ_{2})$.

The projective representation of SO(4) appears in fermionic line operators. On a non-spin manifold $\CM$, we are not allowed to have a chargeless fermion. However, a charged fermion can exist on $\CM$ because it only requires the Spin$^\BC$ structure and any oriented four-manifold admits the Spin$^\BC$ structure~\cite{teichner1994all}. This gives the spin/charge relation and can be understood as the Lorentz symmetry fractionalization.

Consider Maxwell theory and take a subgroup $G^{(1)} = \BZ_2^e\times\BZ_2^m\subset \U(1)_e\times\U(1)_m$ of the electric and magnetic one-form symmetries.
To realize symmetry fractionalization, we turn on the two-form background field $B_e$ and $B_m$ for $G^{(1)} = \BZ_2^e\times\BZ_2^m$ using the background field for the spacetime symmetry group $\SO(4)$~\cite{Thorngren:2014pza,Cordova:2018acb,Wang:2018qoy,Hsin:2019fhf}.
We consider four options to introduce the background fields
\begin{align}
\label{eq:bgf_option}
    (B_e,B_m) = (0,0)\,,\quad \left(0,\pi w_2(\CM)\right)\,,\quad (\pi w_2(\CM),0)\,,\quad (\pi w_2(\CM),\pi w_2(\CM))\,.
\end{align}
Activating the two-form background in this way does not change the spectrum and correlation functions but the statistics of Wilson and 't~Hooft lines.

Each choice attaches a different statistical pattern to the set of line operators.
For example, in the case of $(B_e,B_m) = (\pi w_2(\CM),0)$, the fundamental Wilson line becomes fermionic.
To see this, we have the gauge-invariant Wilson line along a one-cycle $\gamma$
\begin{align}
    W = e^{\i \int_\gamma A}\cdot \exp\left(\i\pi\int_\Sigma w_2(\CM)\right)\,,
\end{align}
where $\Sigma$ is a two-dimensional open surface such that $\partial\Sigma = \gamma$.
The surface term reduces to an SPT phase for the $\SO(2)$ rotation symmetry. Thus, the $\SO(2)$ rotation acts as a projective representation, and $2\pi$-rotation yields $(-1)$, which implies that the Wilson line is fermionic~\cite{Cordova:2018acb}.

Alternatively, we can understand this in another way since the choice of the background fields $(B_e,B_m) = (\pi w_2(\CM),0)$ gives the symmetry
\begin{align}
    \mathrm{Spin}^\BC(4) = \frac{\U(1)\times \mathrm{Spin}(4)}{\BZ_2}\,.
\end{align}
Here, the elements $-1\in \U(1)$ and $(-1)^F\in\mathrm{Spin}(4) $ are identified under the $\BZ_2$ quotient.
Thus, for a Wilson line with an odd charge, the fermion parity acts as $-1$, which suggests that such a Wilson line is fermionic~\cite{Hsin:2018vcg}.
We call it a fermionic Wilson line $W_\mathrm{f}$. The other type of Wilson line is bosonic $W_\mathrm{b}$.
The same terminology will be applied to the 't~Hooft lines.

The statistics of the fundamental Wilson and 't~Hooft operators can completely determine those of the general dyonic operators. Consider the statistics of the dyonic operator $W^nT^m$. 
Such an operator can be understood as a bound state of $n$ fundamental Wilson lines and $m$ fundamental 't~Hooft lines.
The quantum number of the bound state comes from those of the fundamental line operators and an additional contribution by the electromagnetic field~\cite{Jackiw:1976xx,Hasenfratz:1976gr,Goldhaber:1976dp}.
The statistics label $\sigma_d\in\BZ_2$ of the dyonic operator is given by
\begin{align}
\label{eq:dyonic_stat}
    \sigma_d = n\,\sigma_W + m\,\sigma_T + n\, m \quad {\rm mod}\ 2\,,
\end{align}
where $\sigma_W\in\BZ_2$ and $\sigma_T\in\BZ_2$ are the statistics labels of the fundamental Wilson and 't~Hooft lines, respectively. The last term represents the angular momentum of the electromagnetic field. For example, when the fundamental Wilson and 't~Hooft lines are bosonic, the dyonic operator $W_\mathrm{b}^1T_\mathrm{b}^1$ has the statistics $\sigma_d = 0 + 0 + 1 =1$ (fermionic).

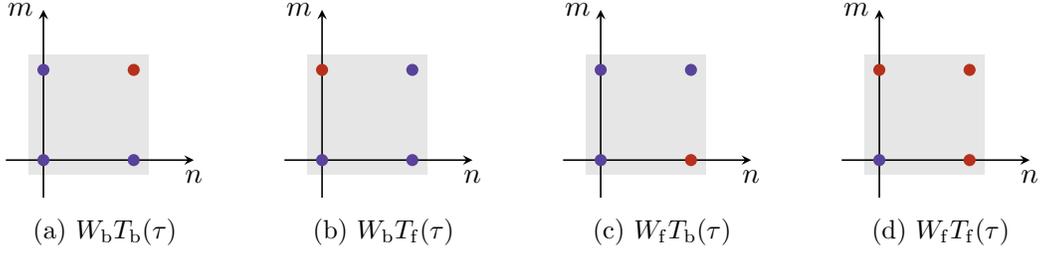
\begin{figure}
    \centering
    \begin{minipage}{0.23\textwidth}
    \begin{center}
        \begin{tikzpicture}[transform shape]
            \fill[gray!20] (-0.2,-0.2)--(1.4,-0.2)--(1.4,1.4)--(-0.2,1.4)--cycle;
            \draw[semithick,->,>=stealth](-0.5,0)--(2,0) node [below] {$n$};
            \draw[semithick,->,>=stealth](0,-0.5)--(0,2) node[left] {$m$};
            \fill[Violet](0,0) circle[radius=0.08cm];
            \fill[Violet](1.2,0) circle[radius=0.08cm];
            \fill[Violet](0,1.2) circle[radius=0.08cm];
            \fill[BrickRed](1.2,1.2) circle[radius=0.08cm];
        \end{tikzpicture}
    \end{center}
    \subcaption{$W_\mathrm{b}T_\mathrm{b}(\tau)$}
    \end{minipage}
    \begin{minipage}{0.23\textwidth}
    \begin{center}
        \begin{tikzpicture}[transform shape]
            \fill[gray!20] (-0.2,-0.2)--(1.4,-0.2)--(1.4,1.4)--(-0.2,1.4)--cycle;
            \draw[semithick,->,>=stealth](-0.5,0)--(2,0) node [below] {$n$};
            \draw[semithick,->,>=stealth](0,-0.5)--(0,2) node[left] {$m$};
            \fill[Violet](0,0) circle[radius=0.08cm];
            \fill[Violet](1.2,0) circle[radius=0.08cm];
            \fill[BrickRed](0,1.2) circle[radius=0.08cm];
            \fill[Violet](1.2,1.2) circle[radius=0.08cm];
        \end{tikzpicture}
    \end{center}
    \subcaption{$W_\mathrm{b}T_\mathrm{f}(\tau)$}
    \end{minipage}
    \begin{minipage}{0.23\textwidth}
    \begin{center}
        \begin{tikzpicture}[transform shape]
            \fill[gray!20] (-0.2,-0.2)--(1.4,-0.2)--(1.4,1.4)--(-0.2,1.4)--cycle;
            \draw[semithick,->,>=stealth](-0.5,0)--(2,0) node [below] {$n$};
            \draw[semithick,->,>=stealth](0,-0.5)--(0,2) node[left] {$m$};
            \fill[Violet](0,0) circle[radius=0.08cm];
            \fill[BrickRed](1.2,0) circle[radius=0.08cm];
            \fill[Violet](0,1.2) circle[radius=0.08cm];
            \fill[Violet](1.2,1.2) circle[radius=0.08cm];
        \end{tikzpicture}
    \end{center}
    \subcaption{$W_\mathrm{f}T_\mathrm{b}(\tau)$}
    \end{minipage}
    \begin{minipage}{0.23\textwidth}
    \begin{center}
        \begin{tikzpicture}[transform shape]
            \fill[gray!20] (-0.2,-0.2)--(1.4,-0.2)--(1.4,1.4)--(-0.2,1.4)--cycle;
            \draw[semithick,->,>=stealth](-0.5,0)--(2,0) node [below] {$n$};
            \draw[semithick,->,>=stealth](0,-0.5)--(0,2) node[left] {$m$};
            \fill[Violet](0,0) circle[radius=0.08cm];
            \fill[BrickRed](1.2,0) circle[radius=0.08cm];
            \fill[BrickRed](0,1.2) circle[radius=0.08cm];
            \fill[BrickRed](1.2,1.2) circle[radius=0.08cm];
        \end{tikzpicture}
    \end{center}
    \subcaption{$W_\mathrm{f}T_\mathrm{f}(\tau)$}
    \end{minipage}
    \caption{The four ways of the Lorentz symmetry fractionalization of Maxwell theory. Each theory is characterized by the statistical pattern of line operators. The horizontal (vertical) axis shows the charge of Wilson ('t~Hooft) lines. Bosonic and fermionic lines are represented by the blue and red dots, respectively.}
    \label{fig:statistics_line}
\end{figure}

We can classify the non-spin Maxwell theory by a pair of statistics of the fundamental Wilson and the fundamental 't~Hooft line.
As a whole, depending on the choices of background gauge fields, we have four Maxwell theories with different statistics of line operators.
We show the statistical pattern of line operators in each theory in Fig.~\ref{fig:statistics_line}.
In the present paper, we call each theory with complex coupling $\tau$ as $W_\mathrm{b}T_\mathrm{b}(\tau)$, $W_\mathrm{b}T_\mathrm{f}(\tau)$, $W_\mathrm{f}T_\mathrm{b}(\tau)$, $W_\mathrm{f}T_\mathrm{f}(\tau)$, respectively.
In the rest of section~\ref{sss:lorentz}, we list the four types of non-spin Maxwell theory.

\paragraph{(a) $W_\mathrm{b}T_\mathrm{b}(\tau)$}
Without turning on background fields, the fundamental Wilson and 't~Hooft lines are bosonic, namely, in a genuine representation of $\SO(4)$. On the other hand, the dyonic operator $W_\mathrm{b}^1T_\mathrm{b}^1$ becomes fermionic due to the quantum number of the electromagnetic field. Its action is
\begin{align}
    S = \frac{1}{2e^2}  \int_\CM F\wedge\hodge F + \frac{\i\,\theta}{8\pi^2} \int_\CM F\wedge F\,.
\end{align}

\paragraph{(b) $W_\mathrm{b}T_\mathrm{f}(\tau)$}
By setting the magnetic background field $B_m = \pi w_2(\CM)$, the fundamental 't~Hooft line is fractionalized under $\SO(4)$~\cite{Thorngren:2014pza}. We obtain the fermionic 't~Hooft line whose action is given by
\begin{align}
    S = \frac{1}{2e^2}\int_\CM  F\wedge \hodge F + \frac{\i\,\theta}{8\pi^2} \int_\CM F\wedge F + \frac{\i}{2\pi} \int_\CM F\wedge\pi w_2(\CM)\,.
\end{align}

\paragraph{(c) $W_\mathrm{f}T_\mathrm{b}(\tau)$}
As discussed previously, by activating the electric background gauge field, we obtain a non-spin Maxwell theory with a fermionic Wilson line. The action is
\begin{align}
\begin{aligned}
    S=\frac{1}{2e^2} \int_\CM \left|F-\pi w_{2}(\CM)\right|^{2}+ \frac{\i\,\theta}{8\pi^2}\int_\CM (F-\pi w_{2}(\CM))^{2}\,.
\end{aligned}
\end{align}

\paragraph{(d) $W_\mathrm{f}T_\mathrm{f}(\tau)$}
Turning on both magnetic and electric one-form background gauge fields gives Maxwell theory $W_\mathrm{f}T_\mathrm{f}$ known as all-fermion electrodynamics~\cite{wang2014classification,Thorngren:2014pza,Kravec:2014aza,Wang:2018qoy}. Its action can be written as 
\begin{align}
\begin{aligned}
    S=\frac{1}{2e^2} \int_\CM \left|F-\pi w_{2}(\CM)\right|^{2}+& \frac{\i\,\theta}{8\pi^2}\int_\CM (F-\pi w_{2}(\CM))^{2} \\
    &+ \frac{\i}{2\pi} \int_\CM (F-\pi w_2(\CM)) \wedge\pi w_2(\CM) \,.
\end{aligned}
\end{align}
As in the discussion below Eq.~\eqref{eq:gauge_transf}, this theory has a purely gravitational anomaly,~\footnote{This gravitational anomaly is characterized by the five-dimensional action
\begin{align}
	\pi\int_{\cal N}w_2({\cal N})\, \cup \beta\left(w_2({\cal N})\right)=\pi \int_{\cal N}w_2({\cal N})\, \cup w_3({\cal N}),
\end{align}
where $\beta$ is the Bockstein homomorphism associated with the short exact sequence $0 \to \BZ_2 \to  \BZ_4 \to \BZ_2 \to 0$.
} while the other three theories are non-anomalous.

\subsubsection{\texorpdfstring{$\mathrm{SL}(2,\BZ)$}{} duality}
\label{sss: sl2z_duality}

Maxwell theory enjoys an equivalence relation between different coupling constants by $\mathrm{SL}(2,\BZ)$ duality.
On a spin manifold $\CM$, the difference between the four versions of Maxwell theory is invisible.
The electric-magnetic duality exchanges a $\U(1)$ gauge field with its dual field~\cite{Witten:1995gf,Gaiotto:2008ak,Kapustin:2009av,Metlitski:2015yqa}, and acts on the coupling constant $\tau$ as
\begin{align}
\label{eq:s_trans}
    \BS:\;\tau \to -1/\tau\,.
\end{align}
In the simplest case $\theta=0$, the $\BS$-transformation yields $e\to 2\pi/e$ and connects between weak and strong couplings.
Additionally, Maxwell theory on a spin manifold has the equivalence
\begin{align}
\label{eq:t_trans}
    \BT:\;\tau \to \tau+ 1\,.
\end{align}
These dualities dictate the equivalence between Maxwell theory with different couplings on a spin manifold.\footnote{The theory has a mixed anomaly between the $\mathrm{SL}(2,\BZ)$ duality and background gravity~\cite{Witten:1995gf,Verlinde:1995mz,Seiberg:2018ntt}. The pure $\mathrm{SL}(2,\BZ)$ anomaly is analyzed in~\cite{Hsieh:2019iba,Hsieh:2020jpj}.}
Since $\BS$ and $\BT$ generate the group $\mathrm{SL}(2,\BZ)$, Maxwell theory with coupling $\tau$ is equivalent to the one with coupling $\tau'=(a\tau+b)/(c\tau+d)$ where $a,b,c,d\in\BZ$ and $ad-bc=1$.
Since $\tau$ takes values in the upper half plane $\BH$, the fundamental region is $\BH/\mathrm{PSL}(2,\BZ)$.

\begin{figure}
    \begin{center}
	\begin{tikzpicture}[transform shape, scale=1.2, >=stealth]
		\draw[->, thick] (-0.4,0.3) arc (225:-45:0.45);
		\draw(0,0)node{$W_{\rm b}T_{\rm b}$};
		\draw(-0.8,0.4)node[above]{$\BS$};
		\draw[<->, thick](-0.5,-0.3)--(-1.2,-1);
		\draw(-1.05,-0.45)node{$\BT$};
        \draw(-1.3,-1.4)node{$W_{\rm b}T_{\rm f}$};
		\draw[<->, thick](-0.5,-1.4)--(0.5,-1.4);
		\draw(0,-1.4)node[above]{$\BS$};
		\draw(0.75,-1.4)node[right]{$W_{\rm f}T_{\rm b}$};
	\draw[->, thick] ($(2.2,-1.7)+(-0.3,0)$) arc (-135:135:0.45);
		\draw(2.2,-0.7)node{$\BT$};

  \begin{scope}[xshift=3.5cm, yshift=0.4cm]
      \draw(1,-1.4)node[right]{$W_{\rm f}T_{\rm f}$};
	\draw[->, thick] ($(2.3,-1.7)+(-0.2,0)$) arc (-135:135:0.45);
		\draw(2.2,-0.7)node{$\BS,\BT$};
  \end{scope}
    \end{tikzpicture}
    \end{center}
 
    \caption{Duality map between four types of Maxwell theory on a non-spin manifold under $\mathrm{SL}(2,\BZ)$ generators $\BS$ and $\BT$. The three non-anomalous theories covariantly transform each other, while the anomalous theory remains invariant under the $\mathrm{SL}(2,\BZ)$ transformation.}
    \label{fig:duality}
    \end{figure}
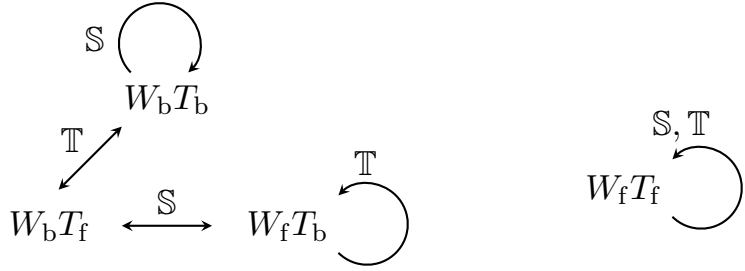

On a non-spin manifold $\CM$, there are four theories that differ in the statistics of their line operators.
Then, it is necessary to follow the duality map between line operators.
The dyonic line $W^nT^m$ is subject to the transformation law~\cite{Witten:1995gf,Metlitski:2015yqa,Ang:2019txy}
\begin{align}
\label{eq:nonspin_line_sl2z}
    \BS:\; W^nT^m\to W^{m} T^{-n}\,,\qquad \BT:\; W^nT^m\to W^{n-m} T^m\,.
\end{align}
Note that the statistics of the line operators remain invariant after the transformation.
Hence, the $\mathrm{SL}(2,\BZ)$ transformation flows a non-spin Maxwell theory to another one:
\begin{align}
    \begin{aligned}
        \BS:\quad W_\mathrm{b}T_\mathrm{b}(\tau)\to W_\mathrm{b} T_\mathrm{b}(-1/\tau)\,,\quad W_\mathrm{b}T_\mathrm{f}(\tau)&\to W_\mathrm{f} T_\mathrm{b}(-1/\tau)\,,\\
        W_\mathrm{f}T_\mathrm{b}(\tau)\to W_\mathrm{b} T_\mathrm{f}(-1/\tau)\,,\quad W_\mathrm{f}T_\mathrm{f}(\tau)&\to W_\mathrm{f} T_\mathrm{f}(-1/\tau)\,,
    \end{aligned}
\end{align}
\begin{align}
    \begin{aligned}
        \BT:\quad W_\mathrm{b}T_\mathrm{b}(\tau)\to W_\mathrm{b} T_\mathrm{f}(\tau+1)\,,\quad W_\mathrm{b}T_\mathrm{f}(\tau)&\to W_\mathrm{b} T_\mathrm{b}(\tau+1)\,,\\
        W_\mathrm{f}T_\mathrm{b}(\tau)\to W_\mathrm{f} T_\mathrm{b}(\tau+1)\,,\quad W_\mathrm{f}T_\mathrm{f}(\tau)&\to W_\mathrm{f} T_\mathrm{f}(\tau+1)\,.
    \end{aligned}
\end{align}
We can summarize the transformation rule as in Fig.~\ref{fig:duality}.
The three non-anomalous theories covariantly transform each other under the $\mathrm{SL}(2,\BZ)$ map, while the anomalous theory $W_\mathrm{f}T_\mathrm{f}(\tau)$ is invariant by itself.\footnote{The transformation rule is analogous with the modular transformation of torus partition functions for 2d fermionic theories. Note that the spin structure $\sigma(a)$ of one-cycle $a\in H_1(\Sigma,\BZ_2)$ obeys $\sigma(a+b) = \sigma(a)+\sigma(b)+\int a\cup b$ where the last term denotes the intersection number between $a$ and $b$~\cite{atiyah1971riemann,Karch:2019lnn}. This is in parallel with the composition rule \eqref{eq:dyonic_stat} fixing the statistical pattern of line operators in Fig.~\ref{fig:statistics_line}. Therefore, each choice of statistical pattern for the line operators is equivalent to a choice of spin structures on a torus.}

\subsection{Topological \texorpdfstring{$\BZ_n$}{} gauge theory}
\label{ss:BF_theory}

Let us consider a four-dimensional pure $\BZ_n$ gauge theory called the BF theory~\cite{Horowitz:1989ng,Maldacena:2001ss,Banks:2010zn}.
The degrees of freedom are a one-form $\U(1)$ gauge field $A$ and a two-form $\U(1)$ gauge field $B$.
The action of the theory is
\begin{align}
\label{eq:BF_action}
    S = \frac{\i n}{2\pi} \int_\CM B\wedge \d A\,,
\end{align}
where $n$ is an integer.
For a while, we assume that our spacetime $\CM$ is closed.
Then, it is invariant under two $\mathrm{U}(1)$ gauge symmetries
\begin{align}
\label{eq:BF_gauge_tf}
    A\to A + \d \lambda^{(0)}\,,\qquad B\to B + \d \lambda^{(1)}\,,
\end{align}
where $\d\lambda^{(i)}$ is not necessarily exact, but $\int \d\lambda^{(i)}\in2\pi\BZ$.
Note that these gauge fields have the standard periodicity for every 2-cycle $\Sigma_2$ and 3-cycle $\Sigma_3$
\begin{align}
\label{eq:periodicity_BF}
    \int_{\Sigma_2} \frac{\d A}{2\pi}\in\BZ\,,\qquad \int_{\Sigma_3} \frac{\d B}{2\pi}\in\BZ\,.
\end{align}
This is necessary to specify a class of line bundles that we sum over in the path integral.
The equations of motion vanish the two gauge-invariant field strengths
\begin{align}
    \d B = 0\,,\qquad F = \d A =0\,.
\end{align}
Hence, no local degrees of freedom appear and the theory turns out to be topological.
Furthermore, from \eqref{eq:periodicity_BF}, the path integral for the gauge fields gives the periodicity condition
\begin{align}
    \int_{\gamma} \frac{A}{2\pi} \in \frac{\BZ}{n}\,,\qquad \int_\Sigma \frac{B}{2\pi}\in\frac{\BZ}{n}\,,
\end{align}
where $\gamma$ is a closed curve and $\Sigma$ is a closed surface.
Equivalently, we can write down $nA = \d \phi$ and $nB=\d \chi$ where $\oint\d\phi \in 2\pi \BZ$ and $\oint\d\chi\in 2\pi\BZ$.
This implies that these fields $A,B$ effectively play the role of $\BZ_n$ gauge fields.

Furthermore, the system has higher-form global symmetries. One is the shift symmetry of the gauge field $A$, and the other is the shift symmetry of $B$:
\begin{align}
\label{eq:BF_global}
    A\to A + \frac{1}{n}\epsilon^{(1)}\,,\qquad B\to B + \frac{1}{n}\epsilon^{(2)}\,,
\end{align}
where $\epsilon^{(i)}$ $(i=1,2)$ is a closed form and $\int \epsilon^{(i)}\in2\pi\BZ$.
The one-form $\BZ_n$ symmetry is generated by a Wilson surface and the two-form global $\BZ_n$ symmetry is generated by a Wilson line.
For a closed path $\gamma$ and a closed surface $\Sigma$ on $\CM$, the Wilson operators are defined as
\begin{align}
    W(\gamma) = e^{\i\oint_\gamma A}\,,\qquad W(\Sigma) = e^{\i\int_\Sigma B}\,.
\end{align}
It is easy to see that these are gauge-invariant under the transformation \eqref{eq:BF_gauge_tf}.
Note that there are no additional 't~Hooft operators except when our spacetime $\CM$ has torsion one-cycle~\cite{Kapustin:2014gua}.
Under the global symmetries~\eqref{eq:BF_global}, these gauge-invariant Wilson operators transform as
\begin{align}
\begin{aligned}
    W(\gamma) &\to e^{\frac{\i}{n}\oint_\gamma \epsilon^{(1)}}\, W(\gamma)\,,\\
    W(\Sigma) &\to e^{\frac{\i}{n}\int_\Sigma \epsilon^{(2)}}\, W(\Sigma)\,.
\end{aligned}
\end{align}
Therefore, for topologically nontrivial cycles $\gamma$ and $\Sigma$, the Wilson operators transform by an $n$-th root of unity. 
These are the fundamental Wilson objects. We can construct a $k$-charged Wilson line $W^k(\gamma)$ and Wilson surface $W^k(\Sigma)$ by composition of $k$ fundamental objects.

\paragraph{Electric coupling to Maxwell theory.}

To illustrate the gauging of the electric $\BZ_n$ symmetry, consider the electric coupling of the BF theory to Maxwell theory
\begin{align}
\label{eq:ele_coulpling_bf_zn}
    S = \frac{1}{2e^2}\int_\CM |F-B|^2 + \frac{\i\,\theta}{8\pi^2} \int_\CM (F-B)^2 + \frac{\i\,n}{2\pi}\int_\CM B\wedge F'\,,
\end{align}
where $F' = \d A'$ is the field strength, and the $B$ field is now a dynamical gauge field.
Here, the gauge transformations are
\begin{align}
    A \to A +  \d\lambda_{\mathrm{em}} + \lambda_e\,,\quad B \to B +\d \lambda_e\,,\quad A' \to A' + \d \lambda_\mathrm{b}\,.
\end{align}
Furthermore, the dual gauge field $\hat{A}'$ for the gauge field $A'$ transforms into $\hat{A}' \to \hat{A}' + \d \lambda_0 + n\lambda_e$ under the gauge transformation (see~\cite{Kapustin:2014gua} for details).
Now we focus on the two one-form global symmetries in the BF theory~\cite{Gaiotto:2014kfa}: One is the one-form symmetry with the current $j_e = nB/2\pi$, which is conserved due to the equation of motion for $A'$. 
Its gauge-invariant charged object is a line operator $\exp(\i \int_\gamma A')$.
After coupling the BF theory to Maxwell theory, this symmetry becomes a new magnetic symmetry combined with the initial magnetic symmetry, and the line operator $\exp(\i \int_\gamma A')$ can be interpreted as the 't~Hooft line.
The other is the one-form symmetry with the current $J_e = F'/2\pi$, which is conserved due to the Bianchi identity.
Naively, its charged objects are line operators consisting of $\hat{A}'$, but this is not gauge invariant.
In the BF theory alone, it is impossible to render this operator gauge-invariant in a non-trivial way.
On the other hand, by using the field in Maxwell theory, the gauge-invariant line operator is constructed as $\exp(\i \int_\gamma (nA-\hat{A}'))$.
This operator is the Wilson loop that is neutral under the electric $\BZ_{n}$ symmetry before gauging.

If one turns on both of the associated background gauge fields, then we have the action
\begin{align}
\label{eq:el_tot}
    S = \frac{1}{2e^2}\int_\CM |F-B|^2 + \frac{\i\,\theta}{8\pi^2} \int_\CM (F-B)^2 + S_\mathrm{BF}[C_1,C_2]\,,
\end{align}
and the BF theory with the background fields $(C_1,C_2)$ is
\begin{align}
\label{eq:bf_bkg}
    S_\mathrm{BF}[C_1,C_2] = \frac{\i\,n}{2\pi}\int_\CM B\wedge (F'-C_1) + \frac{\i}{2\pi} \int_\CM C_2\wedge F'\,,
\end{align}
where $C_i$ $(i=1,2)$ are the two-form $\BZ_n$ gauge fields satisfying $nC_i = \d K_i$ where $\oint\d K_i\in2\pi \BZ$.
Here, the gauge transformations are 
\begin{align}
\begin{aligned}
    A' &\to A'  + \d \lambda_\mathrm{b} + \Lambda_1\,,\quad  C_1 \to C_1 + \d \Lambda_1\,,\\
    B &\to B + \d \lambda_e \,, \qquad \quad\;\, C_2\to C_2 +  \d \Lambda_2\,,
\end{aligned}
\end{align}
in addition to $A\to A +  \d\lambda_{\mathrm{em}} + \lambda_e$.
When both background fields $C_1$, $C_2$ are activated simultaneously, the action~\eqref{eq:el_tot} is not invariant under the gauge transformations due to a mixed anomaly.

Focusing on the $\BZ_2$ symmetry $(n=2)$, one can take the following choices on a non-spin manifold $\CM$:
\begin{align}
\label{eq:bf_choice}
    (C_1,C_2) = (0,0)\,,\quad \left(0,\pi w_2(\CM)\right)\,,\quad (\pi w_2(\CM),0)\,,\quad (\pi w_2(\CM),\pi w_2(\CM))\,.
\end{align}
Here, $w_2(\CM)$ represents the second Stiefel--Whitney class of the tangent bundle.
We sometimes abbreviate $w_2(\CM)$ to $w_2$.
Each choice specifies the type of BF theory on a non-spin manifold.
Although these theories make sense only when coupled with Maxwell theory, we call the BF theory $\mathrm{BF}[C_1,C_2]$ for simplicity.
In what follows, we list the BF theories with non-trivial background gauge fields.

When giving the trivial background gauge fields: $C_1 = C_2 = 0$, we obtain the ordinary BF action for the $\BZ_2$ symmetry
\begin{align}
\label{eq:BF_00}
    S_\mathrm{BF}[0,0] = \frac{\i}{\pi}\int_\CM B\wedge\d A'\,.
\end{align}

Setting the background gauge fields by $C_1 = 0$ and $C_2 = \pi w_2(\CM)$, we have the action
\begin{align}
\label{eq:BF_fS}
    S_\mathrm{BF}[0,\pi w_2] = \frac{\i }{2\pi}\int_\CM \left(2B + \pi\, w_2(\CM)\right) \wedge \d A'\,,
\end{align}
On an oriented manifold, integration by parts gives the BF theory with fermionic strings~\cite{Thorngren:2014pza}. 
In the presence of coupling to Maxwell theory, this procedure breaks gauge invariance because $A'$ is no longer a $\BZ_2$ gauge field.

Once we put $(C_1,C_2) = (\pi w_2(\CM),0)$, the action of the BF theory is
\begin{align}
\label{eq:BF_fL}
    S_\mathrm{BF}[\pi w_2,0] = \frac{\i }{\pi}\int_\CM B\wedge \left(\d A' + \pi w_2(\CM)\right)\,.
\end{align}
This action dictates that the gauge field $A'$ is the $\mathrm{Spin}^\BC$ connection and correspondingly the Wilson line becomes fermionic.

Lastly, turning on both background gauge fields results in the action
\begin{align}
    S_\mathrm{BF}[\pi w_2,\pi w_2] = \frac{\i }{\pi}\int_\CM \left(B\wedge \d A' + \frac{\pi}{2} w_2(\CM)\wedge \d A' + B\wedge \pi w_2(\CM)\right)\,.
\end{align}
This theory has the same gravitational anomaly as the one of all-fermion electrodynamics~\cite{Thorngren:2014pza}.

\paragraph{Magnetic coupling to Maxwell theory.}
Similarly, we can discuss the magnetic coupling of the BF theory to Maxwell theory. The action is
\begin{align}
    S = \frac{1}{2e^2}\int_\CM  F\wedge \hodge F + \frac{\i\,\theta}{8\pi^2} \int_\CM F\wedge F + \frac{\i}{2\pi} \int_\CM B\wedge F + \frac{\i\, n}{2\pi}  \int_\CM B\wedge F'\,,
\end{align}
where $F' = \d A'$ is the field strength.
The gauge transformations are
\begin{align}
    A\to A + \d\lambda_{\mathrm{em}}\,,\quad B \to B + \d\lambda_\mathrm{f}\,,\quad A' \to A' + \d \lambda_\mathrm{b}\,,
\end{align}
and the dual gauge field $\hat{A}'$ transforms into $\hat{A}' \to \hat{A}' + \d\lambda_0 + n\lambda_\mathrm{f}$.
In this case, there are the following two one-form symmetries: One is the one-form symmetry with the current $j_m = nB/2\pi$, which is conserved due to the equation of motion for $F'$.
The corresponding charged object is a Wilson line $\exp(\i\int_\gamma A')$.
The other is the one-form symmetry with the current $J_m = F'/2\pi$, which is conserved due to the Bianchi identity. Its gauge-invariant charged object is an 't~Hooft line $\exp(\i \int_\gamma (\hat{A}'-n \tilde{A}))$ where $\tilde{A}$ is the dual gauge field for the original one $A$ in Maxwell theory, whose gauge transformation is $\tilde{A}\to \tilde{A} + \d \tilde{\lambda}_{\mathrm{em}} + \lambda_\mathrm{f}$.

After gauging, the $\mathrm{U}(1)$ gauge field $A$ is replaced by $nA'$ where $A'$ is the ordinary $\mathrm{U}(1)$ gauge field with an integral periodicity. The fundamental Wilson line before gauging is not fundamental after gauging: $\exp{(\i\int_\gamma A)}\to \exp{(\i\, n\int_\gamma A')} $. The new fundamental Wilson line is $\exp{(\i\int_\gamma A')}$, which comes from the BF theory.

Once we introduce the associated two-form background fields, we have the action
\begin{align}
    S = \frac{1}{2e^2}\int_\CM  F\wedge \hodge F + \frac{\i\,\theta}{8\pi^2} \int_\CM F\wedge F + \frac{\i}{2\pi} \int_\CM B_m\wedge F + S_{\mathrm{BF}}[C_1,C_2]\,,
\end{align}
where the last term is given by \eqref{eq:bf_bkg}.
The above action is the same as \eqref{eq:bf_bkg} in the electric coupling.
The gauge transformations are
\begin{align}
    \begin{aligned}
        A' \to A' + \d\lambda_{\mathrm{b}} + \Lambda_1\,,\quad C_1 \to C_1 + \d\Lambda_1\,,\\
        B \to B + \d\lambda_\mathrm{f}\,,\quad\quad\quad\;\; \,C_2 \to C_2 + \d\Lambda_2\,.
    \end{aligned}
\end{align}
As in the electric coupling, we can choose the background gauge fields $(C_1,C_2)$ by~\eqref{eq:bf_choice} on a non-spin manifold.

\section{Symmetry fractionalization map in Maxwell theory}
\label{sec:frac_map}

This section is devoted to gauging the electric and magnetic $\BZ_2$ one-form symmetry in the Maxwell theories on a non-spin manifold.
To define the $\BZ_2$ gauging consistently, our consideration is on the three non-anomalous theories $W_\mathrm{b}T_\mathrm{b}(\tau)$, $W_\mathrm{b}T_\mathrm{f}(\tau)$ and $W_\mathrm{f}T_\mathrm{b}(\tau)$, excluding the anomalous theory $W_\mathrm{f}T_\mathrm{f}(\tau)$.
We construct the map between the non-anomalous theories and identify their interrelationships by coupling an appropriate TQFT to the Maxwell theories.

Let us pick up the two non-anomalous Maxwell theories $\CT$ and $\CT'$, and construct the map $\CF:\,\CT\mapsto \CT'$ by gauging.
By coupling an appropriate TQFT to the theory $\CT$, the theory after gauging can be equivalent to another Maxwell theory $\CT'$. 
More concretely, we couple a topological $\BZ_2$ gauge theory introduced in section~\ref{ss:BF_theory} to the theory $\CT$ and gauge the diagonal $\BZ_2$ symmetry:
\begin{align}
    \CF:\;\CT \;\longmapsto \;\frac{\CT \times_{\rm e,m} \text{BF}[C_1,C_2]}{\BZ_2}\; \cong\; \CT'\,,
\end{align}
where $\times_\mathrm{e}$ and $\times_\mathrm{m}$ denote the electric and magnetic coupling, respectively, and $\rm{BF}[C_1,C_2]$ is given by the action \eqref{eq:bf_bkg} with $n=2$.
The gauging modifies the background gauge field in the original theory $\CT$ and replaces line operators of $\CT$ to those of $\CT'$.
This provides the interrelationships between the different symmetry fractionalizations and is called the symmetry fractionalization map~\cite{Hsin:2019gvb}.

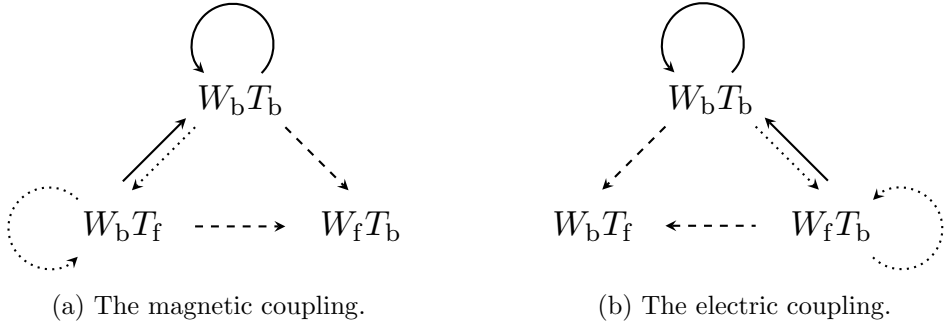
\begin{figure}
\begin{minipage}{0.45\textwidth}
    \begin{center}
	\begin{tikzpicture}[transform shape, >=stealth, scale =1.2]
            \draw(0,0)node{$W_{\rm b}T_{\rm b}$};
            \draw[<-, thick] (-0.4,0.3) arc (225:-45:0.45);
            \draw[->, thick, dashed](0.5,-0.3)--(1.2,-1);
            \draw[->, thick, dotted](-0.5,-0.3)--(-1.2,-1);
            \draw(-1.3,-1.4)node{$W_{\rm b}T_{\rm f}$};
            \draw[->, thick, dotted] (-1.8,-1.1) arc (45:315:0.45);
            \draw[<-, thick](-0.6,-0.2)--(-1.3,-0.9);
            \draw[->, thick, dashed](-0.5,-1.4)--(0.5,-1.4);
            \draw(0.75,-1.4)node[right]{$W_{\rm f}T_{\rm b}$};
        \end{tikzpicture}
    \end{center}
    \subcaption{The magnetic coupling.}
\end{minipage}
\begin{minipage}{0.45\textwidth}
    \begin{center}
	\begin{tikzpicture}[transform shape, >=stealth, scale=1.2]
		\draw(0,0)node{$W_{\rm b}T_{\rm b}$};
            \draw[<-, thick] (-0.4,0.3) arc (225:-45:0.45);
            \draw[->, thick, dashed](-0.5,-0.3)--(-1.2,-1);
            \draw[->, thick, dotted](0.5,-0.3)--(1.2,-1);
            \draw(-1.3,-1.4)node{$W_{\rm b}T_{\rm f}$};
            \draw(0.75,-1.4)node[right]{$W_{\rm f}T_{\rm b}$};
            \draw[<-, thick](0.6,-0.2)--(1.3,-0.9);
            \draw[->, thick, dotted] (1.8,-1.75) arc (-135:135:0.45);
            \draw[<-, thick, dashed](-0.5,-1.4)--(0.5,-1.4);
	\end{tikzpicture}
 \end{center}
        \subcaption{The electric coupling.}
\end{minipage}
    \caption{The symmetry fractionalization maps between the non-spin Maxwell theories by the magnetic gauging (a) and the electric gauging (b). After the $\BZ_2$ gauging, the coupling constant is replaced to $\tau\to 2^2\tau$ in the magnetic case and $\tau\to \tau/2^2$ in the electric case. The type of arrow represents the type of BF action $\mathrm{BF}[C_1,C_2]$. The straight arrow implies the coupling with $\mathrm{BF}[0,0]$ in \eqref{eq:BF_00}. The dashed arrow and the dotted arrow show the coupling with $\mathrm{BF}[\pi w_2,0]$ in \eqref{eq:BF_fL} and the one with $\mathrm{BF}[0,\pi w_2]$ in \eqref{eq:BF_fS}, respectively.}
    \label{fig:fm}
\end{figure}

Before proceeding with the detailed construction of the symmetry fractionalization maps, we summarize our results in Fig.~\ref{fig:fm}.
To construct the symmetry fractionalization maps, we utilize the three BF actions $\mathrm{BF}[0,0]$, $\mathrm{BF}[\pi w_2,0]$, and $\mathrm{BF}[0,\pi w_2]$.
In Fig.~\ref{fig:fm}, they are represented by the straight arrows, the dashed arrows, and the dotted arrows, respectively.
For example, we can read off that the theory $W_\mathrm{b}T_\mathrm{b}$ is mapped to the theory $W_\mathrm{b}T_\mathrm{f}$ when it is electrically coupled to the BF theory $\mathrm{BF}[\pi w_2,0]$:
\begin{align}
    \CF:\;W_\mathrm{b}T_\mathrm{b}(\tau) \;\longmapsto \;\frac{W_\mathrm{b}T_\mathrm{b}(\tau) \times_{\rm e} \text{BF}[\pi w_2,0]}{\BZ_2}\; \cong\; W_\mathrm{b}T_\mathrm{f}(\tau/2^2)\,,
\end{align}
where the coupling constant transforms into $\tau\to \tau/2^2$ by the electric gauging (this map is explicitly constructed in section~\ref{ss:gauging_ele}).
Here, the quotient shows the identification of the $B$ field appearing in Maxwell theory and BF theory. After performing the path integral, the coupled theory turns out to be $W_\mathrm{b}T_\mathrm{f}(\tau/2^2)$ as shown in \eqref{eq:wbtb_w0_ele}.
On the other hand, we cannot obtain $W_\mathrm{b}T_\mathrm{b}$ from $W_\mathrm{b}T_\mathrm{f}$ by the electric gauging due to the mixed anomaly between the electric and magnetic one-form symmetries. In section~\ref{sec:discussion}, we discuss the implications of the maps for quantum dual symmetry. 

In the rest of this section, we show the explicit construction of the symmetry fractionalization maps in Fig.~\ref{fig:fm} by using the path integral formulation.
First, in section~\ref{ss:gauging_mag}, we demonstrate the magnetic gauging to obtain the corresponding symmetry fractionalization maps, as this is somewhat easier than the electric counterpart, and then section~\ref{ss:gauging_ele} focuses on the electric gauging later.

\subsection{Gauging magnetic symmetry}
\label{ss:gauging_mag}
We begin by identifying the symmetry fractionalization map associated with gauging a magnetic $\BZ_{2}$ one-form symmetry with the BF theory.
Note that we can perform the magnetic gauging only for the theories $W_{\mathrm{b}}T_{\mathrm{b}}$ and $W_{\mathrm{b}}T_{\mathrm{f}}$, since this gauging is disturbed by the mixed anomaly in the theory $W_{\mathrm{f}}T_{\mathrm{b}}$.
Hence we have totally six ways of magnetic gauging, as shown in Fig.~\ref{fig:fm}.
In the following, we explicitly demonstrate three of them. 
For the other three cases, we will only make a few comments, but the results are obvious.

\paragraph{$\bm{\mathrm{BF}[C_{1}=0,\,C_{2}=0]}$:} 
We first consider the magnetic coupling $\mathrm{BF}[C_{1}=0,\,C_{2}=0]$ to the theory $W_{\mathrm{b}}T_{\mathrm{b}}(\tau)$.
In this case, there is no subtlety arising from the Stiefel--Whitney $w_{2}(\CM)$.
The corresponding action is given by
\begin{align}
\label{eq:wbtb_00_mag}
    S = \frac{1}{2e^2}\int_\CM  F\wedge \hodge F + \frac{\i\,\theta}{8\pi^2} \int_\CM F\wedge F + \frac{\i}{2\pi} \int_\CM B_m\wedge F + \frac{2\i}{2\pi}\int_\CM B_m\wedge F_m\,.
\end{align}
The equation of motion for $B_{m}$ produces the condition 
\begin{align}
\label{eq:wbtb_00_eom}
F+2F_{m}=0.
\end{align}
This condition restricts the topological sum in the path integral to the sectors, whose period is even, i.e. $\int_{\Sigma}\,F/2\pi\in 2\BZ$ on any closed surface $\Sigma$, which can be regarded as the change of coupling constant $\tau\to 2^{2}\tau$.
Indeed, by replacing $F$ with $-2F_{m}$ in the original theory $W_{\mathrm{b}}T_{\mathrm{b}}(\tau)$, we obtain
\begin{align}
    S = \frac{2^{2}}{2e^2}\int_\CM  F_{m}\wedge \hodge F_{m} + \frac{2^{2}\i\,\theta}{8\pi^2} \int_\CM F_{m}\wedge F_{m}.
\end{align}
Therefore, this gauging procedure gives the symmetry fractionalization map $W_{\mathrm{b}}T_{\mathrm{b}}(\tau)\mapsto W_{\mathrm{b}}T_{\mathrm{b}}(2^{2}\tau)$.

In the case of coupling to the theory $W_{\mathrm{b}}T_{\mathrm{f}}(\tau)$, we have the additional term $\frac{\i}{2\pi}\int_{\CM}F\wedge \pi w_{2}(\CM)$ to the action \eqref{eq:wbtb_00_mag}.
The term does not change the equation of motion for $B_{m}$ \eqref{eq:wbtb_00_eom}, so the replacement $F \to -2F_m$ is still available.
It turns out that the additional term becomes an integer multiple of $2\pi\i$ by the replacement, which is trivial in the exponential form.
As a result, we obtain the map $W_{\mathrm{b}}T_{\mathrm{f}}(\tau)\mapsto W_{\mathrm{b}}T_{\mathrm{b}}(2^{2}\tau)$.

\paragraph{$\bm{\mathrm{BF}[C_{1}=0,\,C_{2}=\pi w_{2}(\CM)]}$:}
We next gauge the magnetic $\BZ_{2}$ symmetry in the theory $W_{\mathrm{b}}T_{\mathrm{f}}(\tau)$ by coupling $\mathrm{BF}[C_{1}=0,\,C_{2}=\pi w_{2}(\CM)]$.
The action of Maxwell theory $W_{\mathrm{b}}T_{\mathrm{f}}(\tau)$ is
\begin{align}
    S = \frac{1}{2e^2}\int_\CM  F\wedge \hodge F + \frac{\i\,\theta}{8\pi^2} \int_\CM F\wedge F &+\frac{\i}{2\pi}\int_{\CM} F\wedge \pi w_{2}(\CM)\,.
\end{align}
We couple this to $\mathrm{BF}[C_{1}=0,\,C_{2}=\pi w_{2}(\CM)]$ by identifying the background field for magnetic one-form symmetry in Maxwell theory and the dynamical $B$ field in the BF theory:
\begin{align}\begin{aligned}
\label{eq:wbtf_0w_mag}
    S = \frac{1}{2e^2}\int_\CM  F\wedge \hodge F + \frac{\i\,\theta}{8\pi^2} \int_\CM F\wedge F &+\frac{\i}{2\pi}\int_{\CM} F\wedge \pi w_{2}(\CM) + \frac{\i}{2\pi} \int_\CM B_m\wedge F\\
    &+\frac{2\i}{2\pi}\int_\CM B_m\wedge F_m +\frac{\i}{2\pi}\int_{\CM}F_{m}\wedge\pi w_{2}(\CM)\,.
\end{aligned}\end{align}
Similar to the previous case, the original $\U(1)$ gauge field $F$ is replaced by $-2F_{m}$ by integrating out $B_{m}$.
The action resulting from this process can be expressed as
\begin{align}
    S = \frac{2^{2}}{2e^2}\int_\CM  F_{m}\wedge \hodge F_{m} + \frac{2^{2}\i\,\theta}{8\pi^2} \int_\CM F_{m}\wedge F_{m} +\frac{\i}{2\pi}\int_{\CM}F_{m}\wedge \pi w_{2}(\CM),
\end{align}
where we have ignored the term $\frac{\i}{2\pi}\int_{\CM}F_{m}\wedge 2\pi w_{2}(\CM)$ because it is an integer multiple of $2\pi\i$.
This action is that of the theory $W_{\mathrm{b}}T_{\mathrm{f}}$, but with the coupling constant different from the original action.
Hence, the symmetry fractionalization map $W_{\mathrm{b}}T_{\mathrm{f}}(\tau)\mapsto W_{\mathrm{b}}T_{\mathrm{f}}(2^{2}\tau)$ is obtained.

Note that in the above argument, the third term in the action \eqref{eq:wbtf_0w_mag}, which makes $W_{\mathrm{b}}T_{\mathrm{b}}$ to be $W_{\mathrm{b}}T_{\mathrm{f}}$, does not contribute to the result of the gauging procedure.
This fact implies that the coupling $\mathrm{BF}[0,\,\pi w_{2}(\CM)]$ to $W_{\mathrm{b}}T_{\mathrm{b}}(\tau)$ yields the map $W_{\mathrm{b}}T_{\mathrm{b}}(\tau)\mapsto W_{\mathrm{b}}T_{\mathrm{f}}(2^{2}\tau)$, as well.

\paragraph{$\bm{\mathrm{BF}[C_{1}=\pi w_{2}(\CM),\,C_{2}=0]}$:}
We demonstrate that the magnetic coupling $\mathrm{BF}[C_{1}=\pi w_{2}(\CM),\,C_{2}=0]$ to the theory $W_{\mathrm{b}}T_{\mathrm{f}}(\tau)$ results in the map $W_{\mathrm{b}}T_{\mathrm{f}}(\tau)\to W_{\mathrm{f}}T_{\mathrm{b}}(2^{2}\tau)$.
To this end, we couple Maxwell theory $W_{\mathrm{b}}T_{\mathrm{f}}(\tau)$ to the theory $\mathrm{BF}[C_{1}=\pi w_{2}(\CM),\,C_{2}=0]$ through the magnetic background field. The corresponding action is
\begin{align}\begin{aligned}
\label{eq:wbtf_w0_mag}
    S = \frac{1}{2e^2}\int_\CM  F\wedge \hodge F &+ \frac{\i\,\theta}{8\pi^2} \int_\CM F\wedge F +\frac{\i}{2\pi}\int_{\CM} F\wedge \pi w_{2}(\CM) + \frac{\i}{2\pi} \int_\CM B_m\wedge F\\
    &+\frac{2\i}{2\pi}\int_\CM B_m\wedge (F_m-\pi w_{2}(\CM))+\frac{2\i}{2\pi}\int_{\CM}\pi w_{2}(\CM)\wedge\pi w_{2}(\CM)\,,
\end{aligned}\end{align}
where the last term is a local counter term depending only on the SW class $w_2({\cal M})$.
Now, the equation of motion for $B_{m}$ is given by
\begin{align}
    F+2(F_{m}-\pi w_{2}(\CM))=0.
\end{align}
By substituting this constraint to remove the original gauge field $F$, the action is written as
\begin{align}\begin{aligned}
    S =&\frac{2^{2}}{2e^2}\int_\CM  \left|F_{m}-\pi w_{2}(\CM)\right|^{2} + \frac{2^{2}\i\,\theta}{8\pi^2} \int_\CM \left(F_{m}-\pi w_{2}(\CM)\right)^{2}\\
    &-\frac{\i}{2\pi}\int_{\CM} F_{m}\wedge 2\pi w_{2}(\CM)+\frac{\i}{2\pi}\int_{\CM} 2\pi w_{2}(\CM)\wedge 2\pi w_{2}(\CM)\,.
\end{aligned}\end{align}
Here, the last two terms are an integer multiple of $2\pi\i$, and this action is nothing but that of the theory $W_{\mathrm{f}}T_{\mathrm{b}}(2^{2}\tau)$.

On the other hand, to describe the map $W_{\mathrm{b}}T_{\mathrm{b}}(\tau)\to W_{\mathrm{f}}T_{\mathrm{b}}(2^{2}\tau)$ by the magnetic coupling of $\mathrm{BF}[\pi w_{2}(\CM),\,0]$, no additional counter terms are required. Thus, the appropriate gauged action is given by
\begin{align}\begin{aligned}
\label{eq:wbtb_w0_mag}
    S = \frac{1}{2e^2}\int_\CM  F\wedge \hodge F + \frac{\i\,\theta}{8\pi^2} \int_\CM F\wedge F  &+ \frac{\i}{2\pi} \int_\CM B_m\wedge F\\
    &+\frac{2\i}{2\pi}\int_\CM B_m\wedge (F_m-\pi w_{2}(\CM))\,.
\end{aligned}\end{align}
It is straightforward to show that this action is equivalent to that of the theory $W_{\mathrm{f}}T_{\mathrm{b}}(2^{2}\tau)$ by integrating $B_{m}$.

\subsection{Gauging electric symmetry}
\label{ss:gauging_ele}
In this subsection, we investigate the symmetry fractionalization maps produced from gauging the electric $\BZ_{2}$ symmetry.
These maps are ones from the theory $W_{\mathrm{b}}T_{\mathrm{b}}$ or $W_{\mathrm{f}}T_{\mathrm{b}}$, since gauging electric $\BZ_{2}$ symmetry in the theory $W_{\mathrm{b}}T_{\mathrm{f}}$ is obstructed by the mixed anomaly.
As in the case of gauging magnetic symmetry, there are three choices of BF theories coupled to a theory.

\paragraph{$\bm{\mathrm{BF}[C_{1}=0,\,C_{2}=0]}$:}
To see the typical effect of gauging electric symmetry, we first consider the electric coupling between the theory $W_{\mathrm{b}}T_{\mathrm{b}}(\tau)$ and $\mathrm{BF}[C_{1}=0,\,C_{2}=0]$.
The corresponding action is obtained by setting $n=2$ in the action \eqref{eq:ele_coulpling_bf_zn}:
\begin{align}
\label{eq:wbtb_00_ele}
    S = \frac{1}{2e^2}\int_\CM |F-B_e|^2 + \frac{\i\,\theta}{8\pi^2} \int_\CM (F-B_e)^2 + \frac{2\i}{2\pi}\int_\CM B_e\wedge F_e\,.
\end{align}
Recall that the invariance under the one-form gauge transformations
\begin{align}
\label{eq:electric_gauge_sym}
    A \to A +  \lambda\,,\quad B_e \to B_e +\d \lambda\,
\end{align}
is imposed as the gauge symmetry.
Following~\cite{Witten:1995gf}, we carefully implement the path integral with respect to $A_e$.
The path integral for $A_e$ consists of a discrete sum over line bundles and a continuous integral over connections on a fixed line bundle. 
For each line bundle ${\cal L}$, we denote a connection on it by $A_e = V_0 + V'$ where $V_0$ is a fixed connection on the line bundle and $V'$ is a globally defined one-form.
The continuous integral with respect to $V'$ reduces to
\begin{align}
    \int \CD V'\; e^{\frac{2\i}{2\pi}\int V'\wedge\, \d B_e} = \delta(\d B_e)\,.
\end{align}
On the other hand, the discrete sum over line bundles yields
\begin{align}
\label{eq:untwisted_be}
    \sum_{{\cal L}\,\in\, H^2(\CM,\BZ)}\; e^{\frac{2\i}{2\pi}\int \d V_0 \,\wedge\, B_e} = \delta\left(\left[\frac{2 B_e}{2\pi}\right]\in\BZ\right)\,.
\end{align}
The first condition tells us that $B_e$ is flat, while the other means the nontrivial period $\left[B_e/2\pi\right]\in \frac{1}{2}\BZ$.
This is an obstruction to gauging $B_e$ to zero by using the gauge symmetry \eqref{eq:electric_gauge_sym} because we can set $B_e=0$ only if $\d B_e=0$ and the period $\left[B_e/2\pi\right]$ is integral.
This tells us that after gauging we have $\mathrm{U(1)}/\BZ_2$ theory with a fractional flux. By redefining $A\to A/2$, this theory can be described by an ordinary $\mathrm{U}(1)$ gauge theory with an integral flux. Then, we obtain the action
\begin{align}
    S = \frac{1}{2\cdot 2^{2}e^2}\int_\CM F\wedge\hodge F + \frac{\i\,\theta}{8\pi^2\cdot 2^{2}}\int_\CM F\wedge F
\end{align}
This is the theory $W_{\mathrm{b}}T_{\mathrm{b}}$ with the coupling constant $\tau/2^{2}$, and the symmetry fractionalization map $W_{\mathrm{b}}T_{\mathrm{b}}(\tau)\to W_{\mathrm{b}}T_{\mathrm{b}}(\tau/2^{2})$ is obtained.

It is straightforward to extend this argument to the electric coupling to the theory $W_{\mathrm{f}}T_{\mathrm{b}}(\tau)$.
The action is given by
\begin{align}\begin{aligned}
\label{eq:wftb_00_ele}
    S &= \frac{1}{2e^2}\int_\CM |F-\pi w_{2}(\CM)-B_e|^2 + \frac{\i\,\theta}{8\pi^2} \int_\CM (F-\pi w_{2}(\CM)-B_e)^2 + \frac{2\i}{2\pi}\int_\CM B_e\wedge F_e\\
    &= \frac{1}{2e^2}\int_\CM |F-B'_e|^2 + \frac{\i\,\theta}{8\pi^2} \int_\CM (F-B'_e)^2 + \frac{2\i}{2\pi}\int_\CM B'_e\wedge F_e-\frac{\i}{2\pi}\int_{\CM} 2\pi w_{2}(\CM)\wedge F_{e}\,,
\end{aligned}\end{align}
where the integrated variable $B_{e}$ is shifted to $B'_{e}:=B_{e}+\pi w_{2}(\CM)$ in order to obtain the second line.
Note that we can ignore the last term because this term is an integer multiple of $2\pi\i$.
Similar to the argument in the previous paragraph, the two-form gauge field $B'_{e}$ is removed after rescaling the one-form gauge field as $A\to A/2$.
The resulting theory is again $W_{b}T_{b}(\tau/2^{2})$.
Therefore, this procedure corresponds to the the map $W_{\mathrm{f}}T_{\mathrm{b}}(\tau)\to W_{\mathrm{b}}T_{\mathrm{b}}(\tau/2^{2})$.

\paragraph{$\bm{\mathrm{BF}[C_{1}=0,\,C_{2}=\pi w_{2}(\CM)]}$:}
Our next aim is the construction of the map $W_{\mathrm{f}}T_{\mathrm{b}}(\tau)\to W_{\mathrm{f}}T_{\mathrm{b}}(\tau/2^{2})$.
We start with the action of Maxwell theory $W_{\mathrm{f}}T_{\mathrm{b}}(\tau)$
\begin{align}
    S = \frac{1}{2e^2}\int_\CM |F-\pi w_{2}(\CM)|^2 &+ \frac{\i\,\theta}{8\pi^2} \int_\CM (F-\pi w_{2}(\CM))^2\,.
\end{align}
Then, we electrically couple $\mathrm{BF}[C_{1}=0,\,C_{2}=\pi w_{2}(\CM)]$ to the theory $W_{\mathrm{f}}T_{\mathrm{b}}(\tau)$:
\begin{align}\begin{aligned}
\label{eq:wftb_0w_ele}
    S = \frac{1}{2e^2}\int_\CM |F-\pi w_{2}(\CM)-B_e|^2 &+ \frac{\i\,\theta}{8\pi^2} \int_\CM (F-\pi w_{2}(\CM)-B_e)^2\\
    &+ \frac{2\i}{2\pi}\int_\CM B_e\wedge F_e+\frac{\i}{2\pi}\int_{\CM} F_{e}\wedge\pi w_{2}(\CM)\,.
\end{aligned}\end{align}
Again, we divide the integral over $A_{e}$ into the discrete sum of line bundles and the continuous integral over the globally defined one-form.
The continuous integral ensures $B_e$ to be flat, and the discrete sum produces the condition
\begin{align}
\label{eq:twisted_be}
    \int_{\Sigma} \frac{2B_{e}}{2\pi}=\frac{1}{2}\int_{\Sigma} w_{2}(\CM)\quad \mathrm{mod}\,\BZ\,,
\end{align}
for any two-cycle $\Sigma$.
Now, in order to eliminate $\pi w_{2}(\CM)$ and $B_{e}$ from the expression $F-\pi w_{2}(\CM)-B_{e}$, we attempt to rescale the original connection as $A\to A/2$.
After this rescaling, the action is written as
\begin{align}
    S = \frac{1}{2\cdot 2^{2}e^2}\int_\CM |F-2\pi w_{2}(\CM)-2B_e|^2 &+ \frac{\i\,\theta}{8\pi^2\cdot 2^{2}} \int_\CM (F-2\pi w_{2}(\CM)-2B_e)^2\,.
\end{align}
The $2\pi w_{2}(\CM)$ part in this expression is removed by the one-form gauge transformation \eqref{eq:electric_gauge_sym}.
However, we cannot eliminate the $B_{e}$ part due to the non-trivial contribution from the right-hand side of the constraint~\eqref{eq:twisted_be}.
This obstruction implies that $F-2B_{e}$ is nothing but the $\mathrm{Spin}^{\BC}$ connection.
Hence, we conclude that this theory is $W_{\mathrm{f}}T_{\mathrm{b}}(\tau/2^{2})$.

The discussion of the theory $W_{\mathrm{b}}T_{\mathrm{b}}(\tau)$ is the same.
We can easily check that the electric coupling of $\mathrm{BF}[0,\,\pi w_{2}(\CM)]$ to $W_{\mathrm{b}}T_{\mathrm{b}}(\tau)$ yields the map $W_{\mathrm{b}}T_{\mathrm{b}}(\tau)\to W_{\mathrm{f}}T_{\mathrm{b}}(\tau/2^{2})$.

\paragraph{$\bm{\mathrm{BF}[C_{1}=\pi w_{2}(\CM),\,C_{2}=0]}$:}
One of the remaining maps is $W_{\mathrm{f}}T_{\mathrm{b}}(\tau)\to W_{\mathrm{b}}T_{\mathrm{f}}(\tau/2^{2})$.
In order to construct this map, we consider coupling $\mathrm{BF}[\pi w_{2}(\CM),\,0]$ to $W_{\mathrm{f}}T_{\mathrm{b}}(\tau)$ with the appropriate counter term.
The action with the counter term is given by
\begin{align}\begin{aligned}
\label{eq:wftb_w0_ele}
    S =&\frac{1}{2e^2}\int_\CM |F-\pi w_{2}(\CM)-B_e|^2 + \frac{\i\,\theta}{8\pi^2} \int_\CM (F-\pi w_{2}(\CM)-B_e)^2\\
    &+ \frac{2\i}{2\pi}\int_\CM B_e\wedge \left(F_e-\pi w_{2}(\CM)\right)+\frac{2\i}{2\pi}\int_{\CM} \pi w_{2}(\CM)\wedge\pi w_{2}(\CM)\,.
\end{aligned}\end{align}
By performing the integral over $A_{e}$, $B_{e}$ is closed and constrained to be $\BZ_{2}$-valued.
Then, it is convenient to define the shifted two-form gauge field as $B'_{e}=B_{e}+\pi w_{2}(\CM)$ and write down the action in terms of this variable.
At this stage, the action is written as
\begin{align}\begin{aligned}
    S =&\frac{1}{2e^2}\int_\CM |F-B'_e|^2 + \frac{\i\,\theta}{8\pi^2} \int_\CM (F-B'_e)^2\\
    &- \frac{\i}{2\pi}\int_\CM 2B'_e\wedge\pi w_{2}(\CM)+\frac{\i}{2\pi}\int_{\CM} 2\pi w_{2}(\CM)\wedge 2\pi w_{2}(\CM)\,.
\end{aligned}\end{align}
The last term is trivial, and the two-from gauge field $B_{e}$ appearing in the first and second term can be removed after the rescaling of $A$.
Naively, the third term also seems to vanish by the one-form gauge transformation \eqref{eq:electric_gauge_sym}.
Recall, however, that we cannot set $B_{e}$ to be zero due to the constraint \eqref{eq:untwisted_be}.
Rather, this obstruction is passed on to the $F$ after the rescaling, and the third term is written as $\frac{\i}{2\pi}\int_{\CM}F\wedge\pi w_{2}(\CM)$.
Therefore, it turns out that this theory is $W_{\mathrm{b}}T_{\mathrm{f}}(\tau/2^{2})$.

The map $W_{\mathrm{b}}T_{\mathrm{b}}(\tau)\to W_{\mathrm{b}}T_{\mathrm{f}}(\tau/2^{2})$ is obtained by the same discussion.
We consider the action
\begin{align}
\label{eq:wbtb_w0_ele}
    S =\frac{1}{2e^2}\int_\CM |F-B_e|^2 + \frac{\i\,\theta}{8\pi^2} \int_\CM (F-B_e)^2+ \frac{2\i}{2\pi}\int_\CM B_e\wedge \left(F_e-\pi w_{2}(\CM)\right)\,.
\end{align}
Note that we do not need the additional counter term in the above action.
By repeating the argument in the previous paragraph, we can easily achieve the theory $W_{\mathrm{b}}T_{\mathrm{f}}(\tau/2^{2})$.

\section{Duality defects of non-spin Maxwell theory}
\label{sec:top_defect}
Maxwell theory defined on a spin manifold has $\mathrm{SL}(2,\BZ)$ duality and one-form symmetries. 
We can combine the action of the $\mathrm{SL}(2,\mathbb{Z})$ transformation with the gauging procedure to define more general operations. These operations map a theory with a given coupling constant to the same theory but with a different coupling constant.
At a particular value of the coupling constant, the theory becomes self-dual under certain operations.
For instance, the operation $\BS$ defined in \eqref{eq:s_trans} maps the theory with $\tau=\i$ to itself.
A simple non-trivial example appears at $\tau=2\i$.
At the coupling constant, the theory is self-dual under the combination of gauging electric $\BZ_{2}$ symmetry and $\BS$-transformation:
\begin{align}
    \tau=2\i \; \xlongrightarrow{\BZ_{2}^e} \; 2\i/2^{2} \; \xlongrightarrow{\BS} \; 2\i.
\end{align}
Such self-dual operations can be interpreted as symmetries of Maxwell theory.
This implies that the topological defects associated with these symmetries can be constructed.
In general, it turns out that this type of symmetry is non-invertible~\cite{Choi:2021kmx,Kaidi:2021xfk,Niro:2022ctq,Cordova:2023ent}. Correspondingly, the algebraic structure of the symmetry is characterized by not a group-like fusion rule but a more general fusion rule, which has no inverse operator.

To construct a general topological defect, it is convenient to prepare two types of topological interfaces. One acts as $\mathrm{SL}(2,\BZ)$ transformation, and the other implements the gauging of a one-form symmetry.
We call these interfaces $\mathrm{SL}(2,\BZ)$ interfaces and gauging interfaces, respectively.
A general topological defect can be composed by stacking these interfaces~\cite{Cordova:2023ent}.

Let us consider Maxwell theories on a non-spin manifold.
As reviewed in subsection \ref{sss: sl2z_duality}, the three non-anomalous theories are related by $\mathrm{SL}(2,\BZ)$ duality.
Besides, we have identified the effect of gauging one-form symmetries in section \ref{sec:frac_map}.
Now, we proceed to investigate topological defects constructed by these structures on a non-spin manifold.
At the self-dual point under a composition of $\mathrm{SL}(2,\BZ)$ and the gauging procedure, there is a corresponding topological defect.
To reveal the nature of the defect, we explicitly specify the defect Lagrangian.
To this end, we construct $\mathrm{SL}(2,\BZ)$ interfaces and gauging interfaces.
Note that, although the bulk theory is formulated on a non-spin manifold, the worldvolume of an interface is a spin manifold, since for any closed oriented three-manifold, the second Stiefel-Whitney class is trivial~\cite{MR0440554,MR1809834}.

\subsection{\texorpdfstring{$\mathrm{SL}(2,\BZ)$}{} interfaces}
\label{ss:sl2z_interface}

This section gives the interface actions of $\mathrm{SL}(2,\BZ)$ duality transformation in non-spin Maxwell theories.
We consider the interface actions of the $\BS$-transformation and the $\BT$-transformation because they are generators of $\mathrm{SL}(2,\BZ)$ transformation.
A general $\mathrm{SL}(2,\BZ)$ interface can be constructed by stacking them together.

In our setup, the spacetime manifold $\CM$ is decomposed into left and right parts as
\begin{align}
    \CM=\CM_{L}\cup\CM_{R},
\end{align}
where we define an interface on the codimension-one submanifold $\partial\CM_{L}=\partial\overline{\CM_R}=W$.
Given the $\U(1)$ gauge fields $A_{L}$ and $A_{R}$ on the left and right, the total Lagrangian with an interface action takes the form
\begin{align}
\label{eq:dl_intf}
    S=\int_{\CM_{L}}\mathcal{L}_{L}(A_{L};\tau)+\int_{W}\mathcal{L}_{S}(A_{L},A_{R})+\int_{\CM_{R}}\mathcal{L}_{R}(A_{R};\tau')\,,
\end{align}
where $\mathcal{L}_{L}$ and $\mathcal{L}_{R}$ are Lagrangians of an appropriate anomaly-free theory and may include $w_{2}$.
The second Stiefel-Whitney class that appears in the interface Lagrangian $\mathcal{L}_{S}$ should originally be represented as $w_{2}(\CM)|_{W}$ or $w_{2}(W)$, but in the following we will simply denote it as $w_{2}$.
On a spin manifold, the duality transformation does not change the theory itself and only changes its coupling constant $\tau\to\tau'$.
On the other hand, the $\mathrm{SL}(2,\BZ)$ transformation for non-spin Maxwell theories alters the type of theory in addition to the coupling constant as in Fig.~\ref{fig:duality}.
In what follows, we construct each $\mathrm{SL}(2,\BZ)$ generator interface depending on the theories living on the left and right parts.

\subsubsection{\texorpdfstring{$\BS$}{}-transformation}
We describe the interface actions of the $\BS$-transformation in non-spin Maxwell theories.
From Fig.~\ref{fig:duality}, we have three types of the $\BS$-transformation interfaces acting as
\begin{align}
    W_{\mathrm{b}}T_{\mathrm{b}}(\tau)\to W_{\mathrm{b}}T_{\mathrm{b}}(-1/\tau),\quad W_{\mathrm{b}}T_{\mathrm{f}}(\tau)\to W_{\mathrm{f}}T_{\mathrm{b}}(-1/\tau),\quad W_{\mathrm{f}}T_{\mathrm{b}}(\tau)\to W_{\mathrm{b}}T_{\mathrm{f}}(-1/\tau)\,.
\end{align}
We often omit the bulk terms and focus only on the interface Lagrangian.

\paragraph{$\bm{W_{\mathrm{b}}T_{\mathrm{b}}\to W_{\mathrm{b}}T_{\mathrm{b}}}$:}
In this case, the action of the $\BS$-transformation interface is given by
\begin{align}
\label{eq:s_inter_bbbb}
    S=\frac{\i}{2\pi}\int_{W}A_{L}\wedge\d A_{R}\,,
\end{align}
and the bulk Lagrangian $\mathcal{L}_{L}$ and $\mathcal{L}_{R}$ are those of the theory $W_{\mathrm{b}}T_{\mathrm{b}}(\tau)$ and $W_{\mathrm{b}}T_{\mathrm{b}}(-1/\tau)$, respectively.
This interface action is the same as the one for Maxwell theories on a spin manifold~\cite{Gaiotto:2008ak,Kapustin:2009av}.
It is obvious that the action is invariant under the gauge transformation $A_L\to A_L+\d\lambda_L, A_R\to A_R + \d\lambda_R$.
The boundary terms in equations of motion for $A_{L}$ and $A_{R}$ yield the following continuity equations on $W$,
\begin{align}
\begin{aligned}
    \frac{1}{e^{2}}\hodge F_{L}+\frac{\i\,\theta}{4\pi^{2}}F_{L}+\frac{\i}{2\pi}F_{R}&=0\,,\\
    -\frac{1}{e'^{2}}\hodge F_{R}-\frac{\i\,\theta'}{4\pi^{2}}F_{R}+\frac{\i}{2\pi}F_{L}&=0\,,
\end{aligned}
\end{align}
where $e'$ and $\theta'$ are the electric coupling and the theta angle in the complex coupling $\tau'=-1/\tau$.
The $\U(1)$ field strength $F_{L}$ and its dual $\tilde{F}_{L}$ are related by
\begin{align}
    \tilde{F}_{L}=-\frac{2\pi\i}{e^{2}}\hodge F_{L}+\frac{\theta}{2\pi}F_{L}\,.
\end{align}
In terms of this dual description, those continuity equations can be written as
\begin{align}\begin{aligned}
\label{eq:s_bbbb_connect}
    F_{R}=-\tilde{F}_{L}\,,\quad 
    \tilde{F}_{R}={F}_{L}\,.
\end{aligned}\end{align}
These equations imply that the gauge fields $A_{L}$ and $A_{R}$ are connected through the $\BS$-transformation.
To verify that the interface $\eqref{eq:s_inter_bbbb}$ is topological, we check that the energy-momentum tensor is continuous when passing through this interface~\cite{Kapustin:2009av}.
The symmetric traceless energy-momentum tensor of the theory $W_{\mathrm{b}}T_{\mathrm{b}}(\tau)$ is given by
\begin{align}
    T_{\mu\nu}=\frac{1}{e^{2}}\left(F_{\mu\alpha}{F^{\alpha}}_{\nu}+\frac{1}{4}\delta_{\mu\nu}F_{\alpha\beta}F^{\alpha\beta}\right)\,,
\end{align}
and a similar expression is valid for the theory $W_{\mathrm{b}}T_{\mathrm{b}}(-1/\tau)$.
Using the matching conditions \eqref{eq:s_bbbb_connect}, we can confirm the continuity of the energy-momentum tensor at the interface $W$.

\paragraph{$\bm{W_{\mathrm{b}}T_{\mathrm{f}}\to W_{\mathrm{f}}T_{\mathrm{b}}}$:}
We next consider the interface of the $\BS$-transformation from $W_{\mathrm{b}}T_{\mathrm{f}}(\tau)$ to $W_{\mathrm{f}}T_{\mathrm{b}}(-1/\tau)$.
The interface action is written as
\begin{align}
\label{eq:s_inter_bffb}
    S=\frac{\i}{2\pi}\int_{W}A_{L}\wedge\left(\d A_{R}-2\pi w_{2}\right)\,.
\end{align}
The action is invariant under the zero-form gauge transformation $A_L\to A_L+\d\lambda_L, A_R\to A_R + \d\lambda_R$, since $\i\int_W \d\lambda_L\wedge w_2$ is an integer multiple of $2\pi \i$.
We reiterate that the electric one-form gauge transformation is given by the first two equations in Eq.~\eqref{eq:gauge_transf} and $W_{\mathrm{f}}T_{\mathrm{b}}$ is defined by the identification of the background two-form gauge field with $w_{2}$.
Consequently, we ensure that the whole action is invariant under the one-form gauge transformation expressed as
\begin{align}
    A_{R}\to A_{R}+\Lambda\,, \quad
    \pi w_{2}\to \pi w_{2}+\d\Lambda\,.
\end{align}
The variation of the interface action under the one-form gauge transformations yields
\begin{align}
    -\frac{\i}{2\pi}\int_{W}A_{L}\wedge \d\Lambda\,,
\end{align}
but this term is canceled by the boundary contribution of the theory $W_{\mathrm{b}}T_{\mathrm{f}}(\tau)$ on $\CM_{L}$.
From the equation of motions, we obtain the matching condition on $W$
\begin{align}
\begin{aligned}
    \frac{1}{e^{2}}\hodge F_{L}+\frac{\i\,\theta}{4\pi^{2}}F_{L}+\frac{\i}{2\pi}\left(F_{R}-\pi w_{2}\right)&=0\,,\\
    -\frac{1}{e'^{2}}\hodge\left(F_{R}-\pi w_{2}\right)-\frac{\i\,\theta'}{4\pi^{2}}\left(F_{R}-\pi w_{2}\right)+\frac{\i}{2\pi}F_{L}&=0\,,
\end{aligned}
\end{align}
where $e'$ and $\theta'$ are again the parameters in the complex coupling $\tau'=-1/\tau$.~\footnote{When the spacetime manifold has no torsion, the expression $\hodge(F-\pi w_{2})$ should be interpreted as follows~\cite{Metlitski:2015yqa}. First, $w_{2}$ is lifted to $H^{2}(\CM,\BZ)$. Then, in the expression $\hodge w_{2}$, the Hodge dual is taken after embedding it into $H^{2}(\CM,\BR)$. A more careful treatment of the coupling between $\U (1)$ gauge fields and discrete gauge fields requires the introduction of differential cohomology.}
As in the previous case, it follows from these matching conditions that the energy-momentum tensor is continuous, and the interface is topological.

\paragraph{$\bm{W_{\mathrm{f}}T_{\mathrm{b}}\to W_{\mathrm{b}}T_{\mathrm{f}}}$:}
The interface action of the $\BS$-transformation that maps the theory $W_{\mathrm{f}}T_{\mathrm{b}}(\tau)$ to $W_{\mathrm{b}}T_{\mathrm{f}}(-1/\tau)$ is given by
\begin{align}
\label{eq:s_inter_fbbf}
    S=\frac{\i}{2\pi}\int_{W}A_{L}\wedge\d A_{R}\,.
\end{align}
Although this interface action is the same as Eq.~\eqref{eq:s_inter_bbbb}, the bulk theories on both sides are different.
The one-form gauge transformation is given by $A_L\to A_L+\Lambda$, and the whole system is gauge invariant in the presence of the interface.
This interface is topological due to the continuity of the energy-momentum tensor.

\subsubsection{\texorpdfstring{$\BT$}{}-transformation}

We construct the interface action implementing the $\BT$-transformation in non-spin Maxwell theories.
From Fig.~\ref{fig:duality}, there are three types of the $\BT$-transformation:
\begin{align}
    W_\mathrm{b}T_\mathrm{b}(\tau) \to W_\mathrm{b}T_\mathrm{f}(\tau+1)\,,\quad
    W_\mathrm{f}T_\mathrm{b}(\tau) \to W_\mathrm{b}T_\mathrm{b}(\tau+1)\,,\quad
    W_\mathrm{f}T_\mathrm{b}(\tau) \to W_\mathrm{f}T_\mathrm{b}(\tau+1)\,.
\end{align}
As in the $\BS$-transformation, we show that our interface actions give appropriate transformation laws and are topological by using the equations of motion on the wall $W$ and the continuity of the energy-momentum tensor on $W$.

\paragraph{$\bm{W_{\mathrm{b}}T_{\mathrm{b}}\to W_{\mathrm{b}}T_{\mathrm{f}}}$:}

Let us consider the interface connecting $W_\mathrm{b}T_\mathrm{b}(\tau)$ to $W_\mathrm{b}T_\mathrm{f}(\tau+1)$.
The interface action is
\begin{align}
    S = \frac{\i}{2\pi} \int_W a\wedge (\d A_L-\d A_R) + \frac{\i}{4\pi}\int_W A_L\wedge \d A_L + \frac{\i}{2\pi}\int_W A_R\wedge \pi w_2\,,
\end{align}
where $a$ is the auxiliary field living only on the wall $W$.
This action is obviously invariant under the gauge transformation $a\to a + \d \phi$ and $A_L \to A_L + \d \lambda_L$.
For the gauge transformation $A_R\to A_R + \d \lambda_R$ and $\pi w_2\to \pi w_2 + \d \Lambda$, the interface action transforms by
\begin{align}
    \delta S = \frac{\i}{2\pi} \int_W \d\lambda_R\wedge \pi w_2 + \frac{\i}{2\pi}\int_W A_R \wedge \d\Lambda\,.
\end{align}
The first term vanishes because $w_2$ is trivial on the wall $W$ and the second term is canceled by the variation from the theory $W_\mathrm{b}T_\mathrm{f}$ on the right region.
Therefore, the interface action is gauge invariant.

To verify its validity, consider the equation of motion on the wall $W$ from \eqref{eq:dl_intf}.
We have the three independent fields on the wall: $A_L$, $A_R$, and the auxiliary field $a$, and their equations of motion are
\begin{align}
    \begin{aligned}
        \frac{1}{e^2}\hodge F_L + \frac{\i\theta}{4\pi^2}F_L + \frac{\i}{2\pi}\d a + \frac{\i}{2\pi} F_L = 0\,,\\
        -\frac{1}{e'^2}\hodge F_R - \frac{\i\theta'}{4\pi^2} F_R - \frac{\i}{2\pi} \d a =0\,,
    \end{aligned}
\end{align}
and $F_L=F_R$. Thus, we can check that the interface action contributes as the $\BT$-transformation $\tau'=\tau+1$.
Also, we can conclude that the interface is topological because the energy-momentum tensor is continuous due to $F_L=F_R$.

\paragraph{$\bm{W_{\mathrm{b}}T_{\mathrm{f}}\to W_{\mathrm{b}}T_{\mathrm{b}}}$:}

For the interface connecting $W_\mathrm{b}T_\mathrm{f}(\tau)$ to $W_\mathrm{b}T_\mathrm{b}(\tau+1)$, its action is 
\begin{align}
    S = \frac{\i}{2\pi} \int_W a\wedge (\d A_L-\d A_R) + \frac{\i}{4\pi}\int_W A_L\wedge \d A_L - \frac{\i}{2\pi}\int_W A_L\wedge \pi w_2\,.
\end{align}
This interface action is invariant under the gauge transformation because of triviality for $w_2$ on $W$.
On the wall $W$, we have the same equations of motion as in the previous case, from which we can check that the interface implements the $\BT$-transformation.

\paragraph{$\bm{W_{\mathrm{f}}T_{\mathrm{b}}\to W_{\mathrm{f}}T_{\mathrm{b}}}$:}
Under the $\BT$-transformation, the statistical pattern of line operators in the theory $W_\mathrm{f}T_\mathrm{b}(\tau)$ remains invariant.
In this case, the interface action is
\begin{align}
    S = \frac{\i}{2\pi} \int_W a\wedge (\d A_L-\d A_R) + \frac{\i}{4\pi}\int_W (A_L-\pi c)\wedge (\d A_L-\pi w_2) \,,
\end{align}
where $c$ is a trivialization of $w_2$ on the wall $W$.
Noting the gauge transformation $\pi c\to\pi c+\Lambda$ from $\pi w_2\to \pi w_2 + \d\Lambda$, the action turns out to be gauge invariant.
This interface action realizes the $\CT$-transformation because the equations of motion on the wall $W$ are
\begin{align}
    \begin{aligned}
        \frac{1}{e^2}\hodge (F_L-\pi w_2) + \frac{\i\theta}{4\pi^2}(F_L-\pi w_2) + \frac{\i}{2\pi}\d a + \frac{\i}{2\pi}(F_L-\pi w_2) = 0\,,\\
        -\frac{1}{e'^2}\hodge (F_R-\pi w_2) - \frac{\i\theta'}{4\pi^2}(F_R-\pi w_2) - \frac{\i}{2\pi}\d a =0\,,
    \end{aligned}
\end{align}
and $F_L = F_R$. It turns out that the interface implements the $\BT$-transformation and is topological because of the continuity of the energy-momentum tensor.

\subsection{Gauging interfaces}
\label{ss:gauging_interface}
Next, we discuss the nature of gauging interfaces.
As discussed in Section~\ref{sec:frac_map}, there are six gauging procedures each for both magnetic and electric coupling, and for each procedure, the corresponding gauging interface can be constructed.
To summarize the results collectively, we introduce a pair of parameters $(\alpha,\beta)$ where $\alpha,\beta\in\{0,1\}$.
We specify the types of the non-anomalous Maxwell theories by
\begin{align}
    (\alpha,\beta) = 
    \begin{dcases}
        (0,0) & \text{for }W_\mathrm{b}T_\mathrm{b}\,,\\
        (0,1) & \text{for }W_\mathrm{b}T_\mathrm{f}\,,\\
        (1,0) & \text{for }W_\mathrm{f}T_\mathrm{b}\,.
    \end{dcases}
\end{align}
The parameters $\alpha$ and $\beta$ reflect the statistics of fundamental Wilson loop $W^{1}$ and fundamental 't~Hooft loop $T^{1}$ in the theory, respectively.
For each parameter, $\alpha,\beta=0$ corresponds to a bosonic line, while $\alpha,\beta=1$ corresponds to a fermionic line.

In terms of this notation, the general action of the magnetic gauging interface between $(\alpha_L,\beta_L)$ and $(\alpha_R,\beta_R)$ is given by
\begin{align}\label{eq:mag_interface}
    S=\frac{\i}{2\pi}\int_{W}a\wedge\left[\d A_{L}-2\left(\d A_{R}-\alpha_{R}\pi w_{2}\right)\right]-\frac{\i}{2\pi}\int_{W}\left(\beta_{L}A_{L}-\beta_{R}A_{R}\right)\wedge \pi w_{2},
\end{align}
where the subscripts $L$ and $R$ denote the theories defined on ${\cal M}_{\rm L}$ and ${\cal M}_{\rm R}$, and $a$ is an additional dynamical $\U(1)$ one-form gauge field defined only on the interface.
The action of the magnetic gauging interface is independent of $\alpha_{\rm L}$, since only $W_{\rm b}T_{\rm b}$ and $W_{\rm b}T_{\rm f}$ can be magnetically gauged.
On the other hand, the general action of the electric gauging interface between $(\alpha_L,\beta_L)$ and $(\alpha_R,\beta_R)$ is given by
\begin{align}\label{eq:ele_interface}
    S=\frac{\i}{2\pi}\int_{W}a\wedge\left[2(\d A_{L}-\alpha_{L}\pi w_{2})-(\d A_{R}-\alpha_{R}\pi w_{2})\right]+\frac{\i}{2\pi}\int_{W}\beta_{R}A_{R}\wedge\pi w_{2},
\end{align}
where the action is independent of $\beta_{\rm L}$ because the theory $W_\mathrm{b}T_\mathrm{f}$ cannot be electrically gauged.

\subsubsection{Half gauging and topological nature}
To ensure that gauging interfaces are topological, we apply the half gauging construction~\cite{Choi:2021kmx}.
As in the construction of the $\mathrm{SL}(2,\BZ)$ interfaces, we decompose the spacetime into left and right regions.
On both regions, we place a single theory whose symmetry we would like to gauge and we connect trivially their $\U(1)$ gauge fields at the interface.
To construct the gauging interface, we couple an appropriate BF theory with the right theory.
At this point, it is necessary to specify a boundary condition for the BF theory at the interface.
Here, we choose to impose the Dirichlet boundary condition for the two-form gauge field
\begin{align}
\label{eq:dirichlet_b}
    \left.B_{m,e}\,\right|_{W}=0\,.
\end{align}
After the path integral on the right side, the resulting interface connects the two theories related by the gauging operation.

Note that the boundary condition~\eqref{eq:dirichlet_b} renders the interface topological for any background field $(C_1,C_2) = (0,0), (\pi w_2,0), (0,\pi w_2)$ in the BF theory.
To understand its topological nature, consider the equation of motion for one-form gauge field $A_{m,e}$ in the BF theory.
For the action $\mathrm{BF}[0,0]$ or $\mathrm{BF}[\pi w_{2},0]$, the equation of motion for $A_{m,e}$ requires that the two-form gauge field $B_{m,e}$ is flat and $\BZ_{2}$-valued.
Thus, there does not occur any effect under the infinitesimal deformations of the locus where the Dirichlet boundary condition~\eqref{eq:dirichlet_b} is imposed.
In the case of $\mathrm{BF}[0,\pi w_{2}]$, the condition for $B_{m,e}$ is replaced to the twisted condition~\eqref{eq:twisted_be}. However, we can infinitesimally deform the position of the interface because the Stiefel--Whitney class $w_2$ becomes trivial on the interface $W$.

Also, we can check that the gauging interfaces are topological for Maxwell theories with a non-trivial fractionalization class.
Since $w_{2}$ is trivial on $W$, the background field in $W_{\mathrm{b}}T_{\mathrm{f}}$ and $W_{\mathrm{f}}T_{\mathrm{b}}$ can be regarded as imposed the Dirichlet boundary condition, which concludes that the interface is topological even when gauging a one-form symmetry in
$W_{\mathrm{b}}T_{\mathrm{f}}$ or $W_{\mathrm{f}}T_{\mathrm{b}}$ on the right region.
For instance, when we obtain the fractionalization map $W_{\mathrm{f}}T_{\mathrm{b}}(\tau)\to W_{\mathrm{b}}T_{\mathrm{b}}(\tau/2^{2})$, two-form gauge field $B_{e}$ in the BF theory is shifted by $B'_{e}=B_{e}+\pi w_{2}$.
In this case, due to the Dirichlet boundary condition for $B_{e}$ and $w_{2}$, the same boundary condition is applied for the new variable $B'_{e}$ and the gauging procedure preserves the topological nature of the interface.

There is another way to verify that the gauging interfaces are topological.
As in the previous subsection, we can use the continuity of the energy-momentum tensor at the interface to ensure the topological nature.
We can check that the energy-momentum tensor is continuous in all cases of the gauging process discussed earlier.

\subsubsection{Gauge invariance and matching conditions}
The whole system must be gauge invariant in the presence of a gauging interface.
We need to care about the zero-form gauge transformations for $A_{L},A_{R}$ and $a$, and their one-form gauge transformations.
These gauge transformations are explicitly given by
\begin{align}\begin{aligned}
    A_{L}&\to A_{L}+\d\lambda_{L}+\alpha_{L}\Lambda\,, &A_{R}&\to A_{R}+\d\lambda_{R}+\alpha_{R}\Lambda\,,\\
    a&\to a+\d\lambda_{W}\,, &\pi w_{2}&\to \pi w_{2}+\d\Lambda\,,
\end{aligned}\end{align}
where $\lambda_{L,R,W}$ are parameters of zero-form gauge transformations, and $\Lambda$ is a parameter of one-form gauge transformation. 
Let us consider the magnetic gauging interface~\eqref{eq:mag_interface}.
The zero-form gauge transformations yield
\begin{align}
    \frac{\i}{2\pi}\int_{W}\d\lambda_{W}\wedge \left[\d A_{L}-2(\d A_{R}-\alpha_{R}\pi w_{2})\right]-\frac{\i}{2\pi}\int_{W}(\beta_{L}\d\lambda_{L}-\beta_{R}\d\lambda_{R})\wedge \pi w_{2}\,.
\end{align}
The first integral is obviously an integer multiple of $2\pi\i$, since $\frac{2\i}{2\pi}\int(A_{R}-\pi w_{2})$ is an integer.
To deal with the second integral, note that $w_{2}$ is trivial on $W$.
In other words, we can treat $w_{2}$ as a cocycle of even number on $W$.
Thus, the second integral is also an integer multiple of $2\pi\i$.
The variation of the interface action~\eqref{eq:mag_interface} under the one-form gauge transformation is
\begin{align}
    -\frac{\i}{2\pi}\int_{W}(\beta_{L}A_{L}-\beta_{R}A_{R})\wedge\d\Lambda\,,
\end{align}
but the boundary terms from the bulk theories exactly cancel this term.
Similarly, we can confirm that the system with the electric gauging interface~\eqref{eq:ele_interface} is also gauge invariant.

To see properties of the interfaces~\eqref{eq:mag_interface} and~\eqref{eq:ele_interface}, we derive the matching conditions between $A_{L}$ and $A_{R}$.
Again, we consider the magnetic gauging interface~\eqref{eq:mag_interface}.
The equation of motion for the additional gauge field $a$ on the interface leads
\begin{align}
\label{eq:mag_sewing}
    F_{L}=2(F_{R}-\alpha_{R}\,\pi w_{2})\,.
\end{align}
This equation implies that the magnetic $\BZ_{2}$ gauging is performed on the interface to rescale the gauge field by $2$.
From the equations of motion for $A_{L}$ and $A_{R}$, we obtain
\begin{align}
\begin{aligned}
    &\frac{\i}{2\pi}\d a+\frac{1}{e^{2}}\hodge F_{L}+\frac{\i\theta}{4\pi^{2}}F_{L}=0\,,\\
    \frac{2\i}{2\pi}\d a+2^{2} &\left[\frac{1}{e^{2}}\hodge(F_{R}- \alpha_{R}\pi w_{2})+\frac{\i\theta}{4\pi^{2}}(F_{R}-\alpha_{R}\pi w_{2})\right]=0\,.
\end{aligned}
\end{align}
As discussed below, these equations will play essential roles in identifying the action of topological defects on line operators.

Likewise, for the electric gauging interface~\eqref{eq:ele_interface}, the equation of motion for $a$ imposes the matching condition
\begin{align}
\label{eq:ele_sewing}
    F_{L}-\alpha_{L}\pi w_{2}=\frac{1}{2}\left(F_{R}-\alpha_{R}\pi w_{2}\right)\,.
\end{align}
It follows from this equation that the gauge field is rescaled by $1/2$ as the result of gauging electric symmetry.
The constraints that connect the bulk fields to $a$ are given by
\begin{gather}
    \frac{2\i}{2\pi}\d a+\frac{1}{e^{2}}\hodge(F_{L}-\alpha_{L}\pi w_{2})+\frac{\i\theta}{4\pi^{2}}(F_{L}-\alpha_{L}\pi w_{2})=0\,,\label{eq:match_ele_L}\\
    \frac{\i}{2\pi}\d a+\frac{1}{2^{2}}\left[\frac{1}{e^{2}}\hodge(F_{R}-\alpha_{R}\pi w_{2})+\frac{\i\theta}{4\pi^{2}}(F_{R}-\alpha_{R}\pi w_{2})\right]=0\,.\label{eq:match_ele_R}
\end{gather}

\subsection{Non-invertible duality defects}
Now we are ready to construct symmetry defects, which connect two identical Maxwell theories with the same coupling constant. 
As mentioned at the outset of this section, the symmetry defects can be constructed by stacking $\mathrm{SL}(2, \BZ)$ interfaces and gauging interfaces.
The Lagrangian of the topological defect is deduced from the fusion of the $\mathrm{SL}(2,\BZ)$ interfaces and the gauging interfaces.
From the defect Lagrangian, we can see the detailed profile of the topological defects.

\subsubsection{Topological defects on non-spin manifold}
In this subsection, we classify the possible topological interfaces made from the $\mathrm{SL}(2,\BZ)$ defects and the gauging defects discussed in the last two subsections.
According to the analysis in~\cite{Niro:2022ctq}, a necessary condition for the defect connecting the identical theories to be topological is
\begin{align}
    \begin{pmatrix}
        F_{L}\\ \i\hodge F_{L}
    \end{pmatrix}=
    K
    \begin{pmatrix}
        F_{R}\\ \i\hodge F_{R}
    \end{pmatrix},\quad
    K=\begin{pmatrix}
        \cos\varphi&\sin\varphi\\
        -\sin\varphi&\cos\varphi
    \end{pmatrix}.
    \end{align}
On a spin manifold, the rotation matrices $K$ that does not change the coupling constant and yields topological defects are restricted to the following cases:\footnote{In terms of the notation in~\cite{Niro:2022ctq}, Case~$2$ corresponds to $N_{L}=0$, and Case~$3$ corresponds to $N_{L}=N_{R}=0$.}

\begin{description}
    \item[Case $1:$] 
    For a rational number $x\in \BQ$ and non-zero rational numbers $y,z\in\BQ$,
    \begin{align}
    \label{eq:case1}
    K=U\cdot G\cdot T:=
    \begin{pmatrix}
        ~1~&~0~\\~y~&~1~
    \end{pmatrix}
    \begin{pmatrix}
        ~z~&~0~\\0&1/z
    \end{pmatrix}
    \begin{pmatrix}
        ~1~&~x~\\~0~&~1~
    \end{pmatrix}\,.
    \end{align}
    
\item[Case $2:$]
For non-zero rational numbers $x,y\in\BQ$, 
\begin{align}
\label{eq:case2}
    K=T'\cdot G'\cdot U':=
    \begin{pmatrix}
        ~1~&x^2/y\\0&1
    \end{pmatrix}
    \begin{pmatrix}
        x/y&0\\~0~&y/x
    \end{pmatrix}
    \begin{pmatrix}
        ~1~&~0~\\-1/y&1
    \end{pmatrix}\,.
\end{align}

\item[Case $3:$]
For non-zero integers $r,s\in\BZ$ such that $\mathrm{gcd}(r,s)=1$,
\begin{align}
\label{eq:case3}
    K=
    \BS^{-1} \cdot G'
    :=\begin{pmatrix}
        ~0~ & ~1~ \\
        -1 & 0
    \end{pmatrix} \begin{pmatrix}
        r/s & 0 \\
        0 & s/r
    \end{pmatrix}\, .
\end{align}

\item[Case $4:$]
\begin{align}
\label{eq:case4}
    K=\begin{pmatrix}
        ~1~&~0~\\~0~&~1~
    \end{pmatrix}\,,\;
    \begin{pmatrix}
        -1&0\\0&-1
    \end{pmatrix}\,.
\end{align}
\end{description}

In the rest of this subsection, we mention the possibility of constructing the topological defects on a non-spin manifold.
Unlike on a spin manifold, there are the three non-anomalous Maxwell theories on a non-spin manifold.
Since an $\mathrm{SL}(2,\BZ)$ operation may map a theory to a different one as in Fig.~\ref{fig:duality}, not all topological defects on a spin manifold can be realized on a non-spin manifold. 

For a given rotation matrix $K$, the corresponding operation preserving the coupling constant ensures the self-duality, since Maxwell theory is determined only by its coupling constant on a spin manifold.
On the other hand, we are required to fix coupling constants as well as their statistics to specify a theory on a non-spin manifold.
Thus, the above rotation matrices $K$ do not guarantee the self-duality of a non-spin theory.
We additionally ask the rotation matrix $K$ not to change the statistics of a theory.

As we will see explicitly, the rotation matrix $K$ can be composed of stacking the duality transformation and gauging interfaces.
In section~\ref{ss:gauging_interface}, we constructed topological interfaces of the magnetic gauging in $W_{\rm b}T_{\rm b}$ and $W_{\rm b}T_{\rm f}$ and the electric gauging in $W_{\rm b}T_{\rm b}$ and $W_{\rm f}T_{\rm b}$.
The remaining gauging interfaces are for the magnetic gauging in $W_{\rm f}T_{\rm b}$ and the electric gauging in $W_{\rm b}T_{\rm f}$, but they are the orientation reversal of the defects we have already constructed.
For instance, the electric gauging from $W_{\rm b}T_{\rm f}(\tau)$ to $W_{\rm b}T_{\rm b}(\tau/2^{2})$ is described by the orientation reversal of the topological interface \eqref{eq:mag_interface} for the magnetic gauging $W_{\rm b}T_{\rm b}(\tau/2^{2})\to W_{\rm b}T_{\rm f}(\tau)$.
This allows us to possess the $\BZ_2$-gauging interfaces gluing two arbitrary non-spin theories.
By appropriately choosing a gauging interface, one can make the statistics of line operators invariant under the action of $K$ when it includes gauging.
Thus, any transformation $K$ with $\BZ_2$-gauging can be realized as a duality defect of a single theory.

Note that any non-invertible defect can be constructed on a non-spin manifold because it always consists of gauging interfaces.
This implies that, as well as on a spin manifold~\cite{Niro:2022ctq}, we can construct an infinite number of duality defects on a non-spin manifold.
The duality defects that do not exist on a non-spin manifold are composed of $\mathrm{SL}(2,Z)$ defects and they are all invertible.

Below we identify the defects that cannot be constructed on a non-spin manifold for each of the four cases.
We focus on the gauging of the $\mathbb{Z}_2$ subgroup of the one-form $\mathrm{U}(1)$ symmetry.

\paragraph{Case $1:$}
In this case, the rotation matrix $K$ is decomposed into three matrices $U$, $G$, and $T$ as in \eqref{eq:case1}.
The coupling constant that is invariant under the rotation $K$ is given by
\begin{align}
    \tau=\frac{1}{2}\left(\frac{1}{y}-\frac{1}{yz^{2}}-x\right)+\i\sqrt{\frac{1}{y^{2}z^{2}}-\frac{1}{4}\left(\frac{1}{y}+\frac{1}{yz^{2}}+x\right)^{2}},
\end{align}
and the rotation angle is related to three parameters as
\begin{align}
    \cos\varphi=\frac{1}{2}\left(z+\frac{1}{z}+xyz\right).
\end{align}
The matrix $G$ represents the gauging of appropriate one-form symmetries.
Since $x$ and $y$ are rational numbers, for some integers $p_{x}$, $q_{x}$, $p_{y}$, and $q_{y}$ subject to $\mathrm{gcd}(p_{x},q_{x})=\mathrm{gcd}(p_{y},q_{y})=1$ and $p_{y},q_{x},q_{y}\neq 0$, they are written as $x=p_x/q_x$ and $y=p_y/q_y$.
For our purpose, it is convenient to further decompose the matrices $U$ and $T$ as
\begin{align}
\begin{aligned}
    U&=\BS^{-1}\cdot G^{-1}\cdot\BT^{p_{y}q_{y}}\cdot G\cdot\BS=
    \begin{pmatrix}
    0 & -1\\
    1 & 0
    \end{pmatrix}
    \begin{pmatrix}
    \frac{1}{q_{y}} & 0\\
    0 & q_{y}
    \end{pmatrix}
    \begin{pmatrix}
    1& -p_{y}q_{y}\\
    0& 1
    \end{pmatrix}
    \begin{pmatrix}
    q_{y}& 0\\
    0& \frac{1}{q_{y}}
    \end{pmatrix}
    \begin{pmatrix}
        0 & 1\\
        -1 & 0
    \end{pmatrix}\,,\\

    T&=G^{-1}\cdot\BT^{-p_{x}q_{x}}\cdot G=
    \begin{pmatrix}
    \frac{1}{q_{x}} & 0\\
    0 & q_{x}
    \end{pmatrix}
    \begin{pmatrix}
    1& -(-p_{x}q_{x})\\
    0& 1
    \end{pmatrix}
    \begin{pmatrix}
    q_{x}& 0\\
    0& \frac{1}{q_{x}}
    \end{pmatrix}\,.
    \end{aligned}
\end{align}
From these expressions, it turns out that the transformations $U$ and $T$ consist of
the $\BS$-transformation, $\BT$-transformation, and the gauging operations.

Our aim is to search for defects that cannot be realized on a non-spin manifold.
To ensure that $K$ does not include the gauging, we set $z = \pm 1$ and $q_{x} = q_{y} = 1$.
Consequently, $K$ reduces to $K = \pm\BS^{-1} \cdot \BT^{p_{y}} \cdot \BS \cdot \BT^{-p_{x}}$.
The properties of $K$ depend on whether $p_{x}$ and $p_{y}$ are even or odd. 
Under the rotation matrix $K$, non-spin Maxwell theories transform as
\begin{align}
\begin{aligned}
    \left(W_\mathrm{b}T_\mathrm{b},\,W_\mathrm{b}T_\mathrm{f},\,W_\mathrm{f}T_\mathrm{b}\right) &\mapsto \left(W_\mathrm{b}T_\mathrm{b},\,W_\mathrm{b}T_\mathrm{f},\,W_\mathrm{f}T_\mathrm{b}\right) \,,\qquad  (p_x,p_y)=(\mathrm{even},\mathrm{even}) \,,\\

    \left(W_\mathrm{b}T_\mathrm{b},\,W_\mathrm{b}T_\mathrm{f},\,W_\mathrm{f}T_\mathrm{b}\right) &\mapsto \left(W_\mathrm{f}T_\mathrm{b},\,W_\mathrm{b}T_\mathrm{f},\,W_\mathrm{b}T_\mathrm{b}\right) \,,\qquad  (p_x,p_y)=(\mathrm{even},\mathrm{odd}) \,,\\

    \left(W_\mathrm{b}T_\mathrm{b},\,W_\mathrm{b}T_\mathrm{f},\,W_\mathrm{f}T_\mathrm{b}\right) &\mapsto \left(W_\mathrm{b}T_\mathrm{f},\,W_\mathrm{b}T_\mathrm{b},\,W_\mathrm{f}T_\mathrm{b}\right) \,,\qquad  (p_x,p_y)=(\mathrm{odd},\mathrm{even}) \,,\\

    \left(W_\mathrm{b}T_\mathrm{b},\,W_\mathrm{b}T_\mathrm{f},\,W_\mathrm{f}T_\mathrm{b}\right) &\mapsto \left(W_\mathrm{b}T_\mathrm{f},\,W_\mathrm{f}T_\mathrm{b},\,W_\mathrm{b}T_\mathrm{b}\right) \,,\qquad  (p_x,p_y)=(\mathrm{odd},\mathrm{odd}) \,.
\end{aligned}
\end{align}
For the theory $W_{\rm b}T_{\rm b}$, the transformation $K$ yields a duality defect if and only if both $p_x$ and $p_y$ are even. 
In contrast, the transformation $K$ gives a defect on $W_{\rm b}T_{\rm f}$ if $p_{x}$ is even, and on $W_{\rm f}T_{\rm b}$ if $p_{y}$ is even.
Otherwise, the matrix $K$ corresponds to an interface gluing different non-spin theories.

\paragraph{Case $2:$}
In this case, the rotation matrix $K$ consists of three factors $T'$, $G'$, $U'$.
The associated coupling constant is given by
\begin{align}
    \tau=\frac{y}{2}+\i\sqrt{x^{2}-\frac{y^{2}}{4}}.
\end{align}
In terms of $x$ and $y$, the rotation angle is represented as $\cos\varphi=y/2x$.
The diagonal matrix $G'$ corresponds to the electric and magnetic gauging defects.
If $K$ includes gauging, we can always construct the corresponding defect on a non-spin manifold.
To trivialize the gauging factor $G'$, we focus on $x/y=\pm1$. Then, the rotation matrix becomes
\begin{align}
    K = \pm
    \begin{pmatrix}
        ~1~ & ~y~ \\
        0 & 1
    \end{pmatrix}
    \begin{pmatrix}
        ~1~ & ~0~ \\
        -1/y & 1
    \end{pmatrix}\,.
\end{align}
By setting $y = p/q$ where $p,q \in\BZ$ with $\mathrm{gcd}(p,q)=1$, we can rewrite 
\begin{align}
\begin{aligned}
    \begin{pmatrix}
        ~1~ & ~y~ \\
        0 & 1
    \end{pmatrix} &= 
    \begin{pmatrix}
        1/q & 0 \\
         ~0~ & ~q~
    \end{pmatrix}
    \begin{pmatrix}
        1 & pq \\
        ~0~ & ~1~
    \end{pmatrix}
    \begin{pmatrix}
        ~q~ & ~0~ \\
        0 & 1/q
    \end{pmatrix}\,,\\
    
    \begin{pmatrix}
        ~1~ & ~0~ \\
        -1/y & 1
    \end{pmatrix} &= 
    \begin{pmatrix}
        ~0~ & ~1~ \\
        -1 & 0
    \end{pmatrix}
    \begin{pmatrix}
        1/p & 0 \\
         ~0~ & ~p~
    \end{pmatrix}
    \begin{pmatrix}
        1 & pq \\
        ~0~ & ~1~
    \end{pmatrix}
    \begin{pmatrix}
        ~p~ & ~0~ \\
        0 & 1/p
    \end{pmatrix}
    \begin{pmatrix}
        ~0~ & -1 \\
        1 & 0
    \end{pmatrix}\,.
\end{aligned}
\end{align}
Each contains the diagonal matrices representing the gauging of electric and magnetic one-form symmetries.
To search for defects that do not involve gauging, we further focus on the cases where $p= q=\pm1$ and $p=-q=\pm1$.
Such a defect may not be possible to construct on a non-spin manifold.
Once we fix $p=q=\pm1$, the rotation matrix is
\begin{align}
\label{eq:case2_mat}
    K = \pm \,\BT^{-1} \cdot \BS^{-1}\cdot \BT^{-1}\cdot\BS\,,
\end{align}
where $\BS$ and $\BT$ denote the $\mathrm{SL}(2,\BZ)$ duality transformation.
When $p=-q=\pm 1$, the rotation matrix becomes
\begin{align}
\label{eq:case2_mat_2}
    K = \pm \,\BT \cdot \BS^{-1}\cdot \BT\cdot\BS\,,
\end{align}
These transformations act on a non-spin Maxwell theory as
\begin{align}
    \left(W_\mathrm{b}T_\mathrm{b},\,W_\mathrm{b}T_\mathrm{f},\,W_\mathrm{f}T_\mathrm{b}\right) \mapsto \left(W_\mathrm{b}T_\mathrm{f},\,W_\mathrm{f}T_\mathrm{b},\,W_\mathrm{b}T_\mathrm{b}\right)\,.
\end{align}
Thus, the rotation matrices $K$ given by \eqref{eq:case2_mat} and \eqref{eq:case2_mat_2} are realized by topological interfaces rather than topological symmetry defects on a non-spin manifold.
The other type of defect given by \eqref{eq:case2} can be constructed on a non-spin manifold.

\paragraph{Case $3:$}
The decomposition~\eqref{eq:case3} shows that the corresponding defects are composed of the inverse of $\BS$-transformation defect and the gauging defect of $\BZ_s^e\times \BZ_r^m$ one-form symmetries.
The coupling constant invariant under the transformation $K$ is $\tau = \i {s/r}$.
The non-invertible defects corresponding to these transformations are constructed in~\cite{Choi:2021kmx}.
When $r/s\neq \pm1$, the corresponding topological defect can be constructed in non-spin Maxwell theories since gauging defects can glue between arbitrary theories.
Thus, the defects that cannot be constructed on a non-spin manifold appear only when $r/s=\pm1$.
Then, the rotation matrix $K$ becomes
\begin{align}
    K = 
    \begin{pmatrix}
        0 & -1\\
        ~1~ & ~0~
    \end{pmatrix}\,,\;
    \begin{pmatrix}
        ~0~ & ~1~ \\
        -1 & 0
    \end{pmatrix}\,,
\end{align}
where these correspond to the $\BS$-duality  and $\BS^{-1}$-duality transformation.
From Fig.~\ref{fig:duality}, the only non-spin Maxwell theory $W_\mathrm{b}T_\mathrm{b}$ admits these defects. In the other theories $W_\mathrm{b}T_\mathrm{f}$ and $W_\mathrm{f}T_\mathrm{b}$, these matrices are realized by topological interfaces rather than defects.

\paragraph{Case $4:$}
This case consists of the identity matrix and its negative.
The identity rotation matrix $K = \mathrm{diag}(1,1)$ is realized by the identity defect and the condensation defects.
The transformation $K = \mathrm{diag}(-1,-1)$ is given by the composition of the charge conjugation, the identity defect, and  condensation defects.
Since the condensation defects are constructed by the composition of gauging defects~\cite{Roumpedakis:2022aik}, we can construct them on a non-spin manifold.
Additionally, non-spin Maxwell theories admit the charge conjugation defect $C=\BS^2$ since any theory is self-dual under $\BS^2$-transformation as in Fig.~\ref{fig:duality}.
Thus, all topological defects in this case can be realized on a non-spin manifold.

In summary, the duality defects that cannot be constructed on a non-spin manifold are listed as follows:
\begin{itemize}
    \item For the theory $W_\mathrm{b}T_\mathrm{b}$, 
    \begin{align}
    \begin{aligned}
        \pm \,\BS^{-1} \cdot \BT^{p_{y}} \cdot \BS \cdot \BT^{-p_{x}}\,,\quad
        \pm \,\BT^{-1} \cdot \BS^{-1}\cdot \BT^{-1}\cdot\BS\,,\quad \pm \,\BT \cdot \BS^{-1}\cdot \BT\cdot\BS\,,
    \end{aligned}
    \end{align}
    where $(p_x,p_y) = (\mathrm{even},\mathrm{odd})\,,\;(\mathrm{odd},\mathrm{even})\,,\;(\mathrm{odd},\mathrm{odd})$.
    \item For the theory $W_\mathrm{b}T_\mathrm{f}$,
        \begin{align}
    \begin{aligned}
        \pm \,\BS^{-1} \cdot \BT^{p_{y}} \cdot \BS \cdot \BT^{-p_{x}}\,,\quad
        \pm \,\BT^{-1} \cdot \BS^{-1}\cdot \BT^{-1}\cdot\BS\,,\quad \pm \,\BT \cdot \BS^{-1}\cdot \BT\cdot\BS\,,\quad \BS^{\pm 1}
    \end{aligned}
    \end{align}
    where $(p_x,p_y) = (\mathrm{even},\mathrm{even})\,,\; (\mathrm{even},\mathrm{odd})$.
    \item For the theory $W_\mathrm{f}T_\mathrm{b}$,
    \begin{align}
        \pm\, \BS^{-1} \cdot \BT^{p_{y}} \cdot \BS \cdot \BT^{-p_{x}}\,,\quad
        \pm \,\BT \cdot \BS^{-1}\cdot \BT\cdot\BS\,,\quad \pm \, \BT^{-1} \cdot \BS^{-1}\cdot \BT^{-1}\cdot\BS\,,\quad \BS^{\pm 1},
    \end{align}
    where $(p_x,p_y) =(\mathrm{even},\mathrm{even})\,,\;(\mathrm{odd},\mathrm{even})$.
\end{itemize}

\subsubsection{Defects at \texorpdfstring{$\tau=\i$}{}}

This section focuses on Maxwell theories with the coupling constant $\tau=\i$ and gives the defect Lagrangians for each construction.
We illustrate the fusion of topological interfaces and the properties of the resulting defects using several examples.

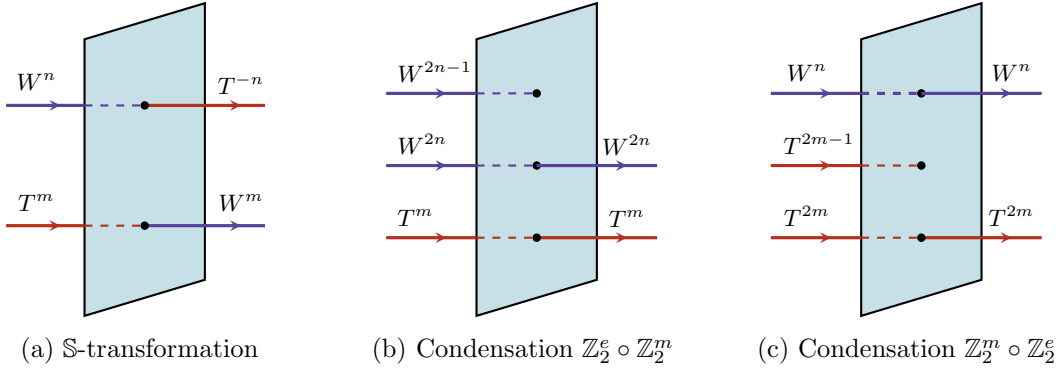
\begin{figure}
    \centering
    \begin{minipage}{0.32\textwidth}
    \begin{center}
        \begin{tikzpicture}[scale=0.8]
          \begin{scope}[yshift=0.1cm]
       \draw[thick, fill=defectblue] (-1,2)--(1,2.6)--(1,-2)--(-1,-2.6)--cycle;
          \end{scope}
	   \coordinate (Wilson) at (-2.3,1);
	   \coordinate (Hooft) at (-2.3,-1);
	   \draw[very thick, Violet] (Wilson)--($(Wilson)+(1.3,0)$);
          \draw[->,>=stealth,thick,Violet] (Wilson)--($(Wilson)+(0.9,0)$);
	   \draw[thick, Violet, dashed] ($(Wilson)+(1.3,0)$)--($(Wilson)+(2.3,0)$);
	   \draw[very thick, BrickRed] ($(Wilson)+(2.3,0)$)--($(Wilson)+(4.3,0)$);
          \draw[->,>=stealth,thick,BrickRed] ($(Wilson)+(2.3,0)$)--($(Wilson)+(3.9,0)$);
          \coordinate (wt_junction) at ($(Wilson)+(2.3,0)$);
          \fill (wt_junction) circle [radius=2pt];
          \coordinate [label=below:\footnotesize $W^{n}$] (upper_W) at ($(Wilson)+(0.5,0.7)$);
          \coordinate [label=below:\footnotesize $T^{-n}$] (upper_T) at ($(Wilson)+(3.9,0.7)$);
	   \draw[very thick, BrickRed] (Hooft)--($(Hooft)+(1.3,0)$);
          \draw[->,>=stealth,thick,BrickRed] (Hooft)--($(Hooft)+(0.9,0)$);
	   \draw[thick, BrickRed, dashed] ($(Hooft)+(1.3,0)$)--($(Hooft)+(2.3,0)$);
	   \draw[very thick, Violet] ($(Hooft)+(2.3,0)$)--($(Hooft)+(4.3,0)$);
          \draw[->,>=stealth,thick,Violet] ($(Hooft)+(2.3,0)$)--($(Hooft)+(3.9,0)$);
          \coordinate (tw_junction) at ($(Hooft)+(2.3,0)$);
          \fill (tw_junction) circle [radius=2pt];
          \coordinate [label=below:\footnotesize $T^{m}$] (lower_W) at ($(Hooft)+(0.5,0.7)$);
          \coordinate [label=below:\footnotesize $W^{m}$] (lower_T) at ($(Hooft)+(3.9,0.7)$);
        \end{tikzpicture}
    \end{center}
    \subcaption{$\BS$-transformation\label{fig:pierce_S}}
    \end{minipage}
    \begin{minipage}{0.32\textwidth}
    \begin{center}
        \begin{tikzpicture}[scale=0.8]
          \begin{scope}[yshift=0.1cm]
       \draw[thick, fill=defectblue] (-1,2)--(1,2.6)--(1,-2)--(-1,-2.6)--cycle;
          \end{scope}
	   \coordinate (Wilson_odd) at (-2.5,1.2);
	   \coordinate (Wilson_even) at (-2.5,0);
	   \coordinate (Hooft) at (-2.5,-1.2);
	   \draw[very thick, Violet] (Wilson_odd) node[above right,black]{\footnotesize $W^{2n-1}$}--($(Wilson_odd)+(1.5,0)$);
          \draw[->,>=stealth,thick,Violet] (Wilson_odd)--($(Wilson_odd)+(1,0)$);
	   \draw[thick, Violet, dashed] ($(Wilson_odd)+(1.5,0)$)--($(Wilson_odd)+(2.5,0)$);
          \coordinate (we_junction) at ($(Wilson_odd)+(2.5,0)$);
          \fill (we_junction) circle [radius=2pt];
          \coordinate (upper_W) at ($(Wilson_odd)+(0.8,0.6)$);
	   \draw[very thick, Violet] (Wilson_even)node[above right,black]{\footnotesize $W^{2n}$}--($(Wilson_even)+(1.5,0)$);
          \draw[->,>=stealth,thick,Violet] (Wilson_even)--($(Wilson_even)+(1,0)$);
	   \draw[thick, Violet, dashed] ($(Wilson_even)+(1.5,0)$)--($(Wilson_even)+(2.5,0)$);
          \coordinate (ww_junction) at ($(Wilson_even)+(2.5,0)$);
          \fill (ww_junction) circle [radius=2pt];
	   \draw[very thick, Violet] ($(Wilson_even)+(2.5,0)$)--($(Wilson_even)+(4.5,0)$);
          \draw[->,>=stealth,thick,Violet] ($(Wilson_even)+(2.5,0)$)--($(Wilson_even)+(4,0)$)node[above,black]{\footnotesize $W^{2n}$};
          \coordinate (mid_W_l) at ($(Wilson_even)+(0.6,0.6)$);
          \coordinate (mid_W_r) at ($(Wilson_even)+(4,0.6)$);
	   \draw[very thick, BrickRed] (Hooft)node[above right,black]{\footnotesize $T^m$}--($(Hooft)+(1.5,0)$);
          \draw[->,>=stealth,thick,BrickRed] (Hooft)--($(Hooft)+(1,0)$);
	   \draw[thick, BrickRed, dashed] ($(Hooft)+(1.5,0)$)--($(Hooft)+(2.5,0)$);
	   \draw[very thick, BrickRed] ($(Hooft)+(2.5,0)$)--($(Hooft)+(4.5,0)$);
          \draw[->,>=stealth,thick,BrickRed] ($(Hooft)+(2.5,0)$)--($(Hooft)+(4,0)$) node[above,black]{\footnotesize $T^m$};
          \coordinate (tt_junction) at ($(Hooft)+(2.5,0)$);
          \fill (tt_junction) circle [radius=2pt];
          \coordinate (lower_T_l) at ($(Hooft)+(0.6,0.6)$);
          \coordinate (lower_T_r) at ($(Hooft)+(4,0.6)$);
        \end{tikzpicture}
    \end{center}
    \subcaption{Condensation $\BZ_{2}^{e} \circ \BZ_{2}^{m}$}
    \label{fig:pierce_em}
    \end{minipage}
    \begin{minipage}{0.32\textwidth}
    \begin{center}
        \begin{tikzpicture}[scale=0.8]
          \begin{scope}[yshift=0.1cm]
	   \draw[thick, fill=defectblue] (-1,2)--(1,2.6)--(1,-2)--(-1,-2.6)--cycle;
          \end{scope}
	   \coordinate (Wilson) at (-2.5,1.2);
	   \coordinate (Hooft_even) at (-2.5,0);
	   \coordinate (Hooft_odd) at (-2.5,-1.2);
    
	   \draw[very thick, Violet] (Wilson)--($(Wilson)+(1.5,0)$);
          \draw[->,>=stealth,thick,Violet] (Wilson)--($(Wilson)+(1,0)$);
	   \draw[very thick, Violet, dashed] ($(Wilson)+(1.5,0)$)--($(Wilson)+(2.5,0)$);
          \coordinate (ww_junction) at ($(Wilson)+(2.5,0)$);
          \fill (ww_junction) circle [radius=2pt];
	   \draw[very thick, Violet] ($(Wilson)+(2.5,0)$)--($(Wilson)+(4.5,0)$);
          \draw[->,>=stealth,thick,Violet] ($(Wilson)+(2.5,0)$)--($(Wilson)+(4,0)$);
          \coordinate [label=below:\footnotesize $W^{n}$] (upper_W_l) at ($(Wilson)+(0.6,0.65)$);
          \coordinate [label=below:\footnotesize $W^{n}$] (upper_W_r) at ($(Wilson)+(4,0.65)$);

	   \draw[very thick, BrickRed] (Hooft_even)--($(Hooft_even)+(1.5,0)$);
          \draw[->,>=stealth,thick,BrickRed] (Hooft_even)--($(Hooft_even)+(1,0)$);
	   \draw[thick, BrickRed, dashed] ($(Hooft_even)+(1.5,0)$)--($(Hooft_even)+(2.5,0)$);
          \coordinate (te_junction) at ($(Hooft_even)+(2.5,0)$);
          \fill (te_junction) circle [radius=2pt];
          \coordinate [label=below:\footnotesize $T^{2m-1}$] (mid_T_l) at ($(Hooft_even)+(0.8,0.7)$);

	   \draw[very thick, BrickRed] (Hooft_odd)--($(Hooft_odd)+(1.5,0)$);
          \draw[->,>=stealth,thick,BrickRed] (Hooft_odd)--($(Hooft_odd)+(1,0)$);
	   \draw[thick, BrickRed, dashed] ($(Hooft_odd)+(1.5,0)$)--($(Hooft_odd)+(2.5,0)$);
	   \draw[very thick, BrickRed] ($(Hooft_odd)+(2.5,0)$)--($(Hooft_odd)+(4.5,0)$);
          \draw[->,>=stealth,thick,BrickRed] ($(Hooft_odd)+(2.5,0)$)--($(Hooft_odd)+(4,0)$);
          \coordinate (tt_junction) at ($(Hooft_odd)+(2.5,0)$);
          \fill (tt_junction) circle [radius=2pt];
          \coordinate [label=below:\footnotesize $T^{2m}$] (lower_T_l) at ($(Hooft_odd)+(0.6,0.7)$);
          \coordinate [label=below:\footnotesize $T^{2m}$] (lower_T_r) at ($(Hooft_odd)+(4,0.7)$);
        \end{tikzpicture}
    \end{center}
    \subcaption{Condensation $\BZ_{2}^{m}\circ\BZ_{2}^{e}$}
    \label{fig:pierce_me}
    \end{minipage}
    \caption{The action of the defects on line operators. The $\BS$-transformation defect (a) swaps the Wilson and 't~Hooft lines. This behavior is consistent with \eqref{eq:nonspin_line_sl2z}. The condensation defects (b) and (c) project out the Wilson and 't~Hooft lines with odd charges, respectively.}
    \label{fig:pierce}
\end{figure}

\begin{figure}
\begin{center}
    \begin{tikzpicture}[scale=0.6]
        \draw[white] (-5,-4) grid (5,3);
        \draw[->, thick] (-4.5,-3.2)--(4.5,-3.2) node[right]{\large $x$};

        \fill[interyellow, draw=none] (-1.9,-3) rectangle (1.9,3);
        \fill[bulkgray, draw=none] (2.1,-3) rectangle (4.5,3);
        \fill[bulkgray, draw=none] (-4.5,-3) rectangle (-2.1,3);
        
        \draw[very thick] (-2,-3)--(-2,3);
        \draw[very thick] (2,-3)--(2,3);
        \node at (-2,-3.7) {$x=0$};
        \node at (2,-3.7) {$x=\delta x$};

        \begin{scope}[yshift=-0.5cm]
        \node at (-3.5,2.5) {\Large ${\cal T}_L$};
        \node at (0,2.5) {\Large ${\cal T}_I$};
        \node at (3.5,2.5) {\Large ${\cal T}_R$};
        \end{scope}

        \node at (-3.5,0) {\large $A_L$};
        \node at (0,0) {\large $a_I$};
        \node at (3.5,0) {\large $A_R$};

    \end{tikzpicture}\hspace{0.2cm}
    \begin{tikzpicture}[scale=0.6]
        
        \draw[white] (-1,-4) grid (1,3);

        \draw[thick, ->, >=stealth] (-1.5,0)--node[above]{$\delta x\to 0$}(1.5,0);
    \end{tikzpicture}\hspace{0.4cm}
    \begin{tikzpicture}[scale=0.6]
       
         \draw[->, >=stealth, thick] (-3.5,-3.2)--(3.5,-3.2)node[right]{\large $x$};

        \draw[thick] (-0.02,-3)--(-0.02,3);
        \draw[thick] (0.02,-3)--(0.02,3);
        \node at (0, -3.7) { $x=0$};

        \fill[bulkgray, draw=none] (0.1,-3) rectangle (3.5,3);
        \fill[bulkgray, draw=none] (-0.1,-3) rectangle (-3.5,3);

        \begin{scope}[yshift=-0.5cm]
        \node at (-2,2.5) {\Large ${\cal T}_L$};
        \node at (2,2.5) {\Large ${\cal T}_R$};
        \end{scope}
        \node at (-2,0) {\large $A_L$};
        \node at (2,0) {\large $A_R$};
    \end{tikzpicture}
\end{center}
\caption{The fusion of two topological interfaces. While an interface at $x=0$ glues the left theory $\CT_L$ with the intermediate theory $\CT_I$, the other at $x=\delta x$ connects the intermediate $\CT_I$ and the right theory $\CT_R$. By stacking them together $(\delta x\to0)$, we obtain a combined topological interface that glues $\CT_L$ and $\CT_R$.}
\label{fig:slab}
\end{figure}
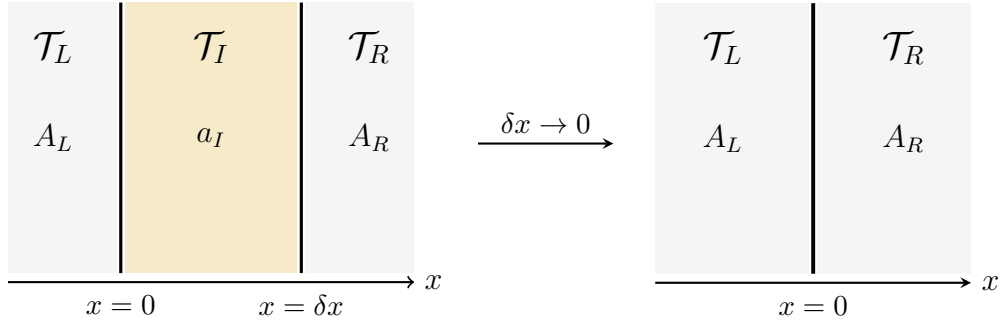

\paragraph{$\bm{W_{\mathrm{b}}T_{\mathrm{b}}(\tau=\i)}$:}
Since the $\BS$-transformation acts as $W_{\mathrm{b}}T_{\mathrm{b}}(\tau)\to W_{\mathrm{b}}T_{\mathrm{b}}(-1/\tau)$, the theory $W_{\mathrm{b}}T_{\mathrm{b}}(\tau=\i)$ is self-dual under this operation.
The Lagrangian of the associated topological defect is given by Eq.~\eqref{eq:s_inter_bbbb}.
The same defect in Maxwell theory on a spin manifold has been constructed in~\cite{Gaiotto:2008ak,Kapustin:2009av}.
Let us consider what happens when a line operator pierces the defect, as in Fig.~\ref{fig:pierce_S}.
Suppose that the defect is penetrated from the left side by the 't~Hooft line with the charge $m$, which is described by the dual field $\tilde{A}_{L}$.
At the junction, the matching condition~\eqref{eq:s_bbbb_connect} is imposed, and the 't~Hooft line is transformed into the Wilson line with the charge $m$.
Similarly, the Wilson line with the charge $n$ from the left side is transformed into the 't~Hooft line with the charge $-n$.
Note that an orientation reversal of this $\BS$-transformation defect implements the inverse operation.
In other words, this $\BS$-transformation defect is invertible.
To see this, we consider fusing of the defect~\eqref{eq:s_inter_bbbb} and its orientation reversal.
When these two defects are inserted, the spacetime is divided into three regions:
    \begin{align}
        \CM=\CM_{L}\cup\CM_{I}\cup\CM_{R}\,.
    \end{align}
We assume that $\CM_{I}$ is a thin slab region and the topology of $\CM_{I}$ is regarded as $W\times I$, where $I=[0,\delta x]$ is an interval.
We denote the local coordinate for the interval $I$ by $x$.
The gauge fields on these three regions are as shown in Fig.~\ref{fig:slab}.
After fusing two defects, the field $a_{I}$ is regarded as the degrees of freedom on the defect, and the resulting Lagrangian is given by
\begin{align}
    S=\frac{\i}{2\pi}\int_{W}A_{L}\wedge \d a_{I}-\frac{\i}{2\pi}\int_{W}a_{I}\wedge \d A_{R}=\frac{\i}{2\pi}\int_{W}\d a_{I}\wedge (A_{L}-A_{R})\,.
\end{align}
This Lagrangian describes the identity defect, since by integrating out the field $a_{I}$, it implies that the fields $A_{L}$ and $A_{R}$ are trivially connected.

We next consider the composition of the two gauging interfaces.
The theory $W_{\mathrm{b}}T_{\mathrm{b}}(\tau=\i)$ is self-dual under the following two operations combined the electric gauging and the magnetic gauging:
\begin{align}
\begin{aligned}
\label{eq:self-dual_bb}
    W_{\mathrm{b}}T_{\mathrm{b}}(\tau=\i) \; \xlongrightarrow{\BZ_{2}^e} \; W_{\mathrm{b}}T_{\mathrm{b}}(\tau=\i/2^{2}) \; \xlongrightarrow{\BZ_{2}^{m}} \; W_{\mathrm{b}}T_{\mathrm{b}}(\tau=\i)\,,\\
    W_{\mathrm{b}}T_{\mathrm{b}}(\tau=\i) \; \xlongrightarrow{\BZ_{2}^e} \; W_{\mathrm{b}}T_{\mathrm{f}}(\tau=\i/2^{2}) \; \xlongrightarrow{\BZ_{2}^{m}} \; W_{\mathrm{b}}T_{\mathrm{b}}(\tau=\i)\,.
\end{aligned}
\end{align}
The corresponding topological defects are constructed by fusing the appropriate gauging defects.
To this end, we again consider the configuration as Fig.~\ref{fig:slab}.
From the expressions for interfaces~\eqref{eq:mag_interface} and~\eqref{eq:ele_interface}, the fused Lagrangian is given by
\begin{align}
\begin{aligned}
    S=&\frac{\i}{2\pi}\int_{W}a_{e}\wedge\left(2\d A_{L}-\d a_{I}\right)+\frac{\i}{2\pi}\int_{W}\beta_{I}a_{I}\wedge\pi w_{2}\\
    &+\frac{\i}{2\pi}\int_{W}a_{m}\wedge \left(\d a_{I}-2\d A_{R}\right)-\frac{\i}{2\pi}\int_{W}\beta_{I}a_{I}\wedge\pi w_{2}\,,
\end{aligned}
\end{align}
where $\beta_{I}=0,1$ for the intermediate theory $\CT_I = W_\mathrm{b}T_\mathrm{b},W_\mathrm{b}T_\mathrm{f}$, and $a_{e,m,I}$ are one-form gauge fields on the electric gauging interface, the magnetic gauging interface, and the slab region, respectively.
Integrating out $a_{I}$ imposes the condition $a:=a_{e}=a_{m}$ and yields
\begin{align}
\label{eq:cond_em_bb}
    S=\frac{2\i}{2\pi}\int_{W}a\wedge(\d A_{L}-\d A_{R})\,.
\end{align}
Although this Lagrangian is similar to one for the identity defect, this defect acts non-trivially on the line operators.
To understand its property, we regard this defect as the stuck of the gauging interface and the electric interface, and consider the Wilson line with charge $n$, penetrating the defect from the left side.
Recall that at the junction with the magnetic interface, the matching condition~\eqref{eq:match_ele_L} is imposed.
In the present case, the matching condition is written as
\begin{align}
    \frac{2\i}{2\pi}\d a+\frac{1}{2\pi}\hodge F_{L}=0.
\end{align}
It follows from this matching condition that the electric charge of the Wilson line measured on a two-sphere $\Sigma$ in the magnetic interface is related to the flux of $a$ as
\begin{align}
    n=-\frac{\i}{2\pi}\int_{\Sigma}\hodge F_{L}=-\frac{2}{2\pi}\int_{\Sigma}\d a\in 2\BZ.
\end{align}
This constraint implies that the Wilson line is projected unless its charge $n$ is even.
See Fig.~\ref{fig:pierce_em} for the summary of the action on line operators.
This type of topological defect is called condensation defect~\cite{Choi:2022zal,Roumpedakis:2022aik}.

Moreover, $W_{\mathrm{b}}T_{\mathrm{b}}(\tau=i)$ exhibits self-duality under the operation that is interchanged the order in the  previous ones~\eqref{eq:self-dual_bb}:
\begin{align}
\begin{aligned}
    W_{\mathrm{b}}T_{\mathrm{b}}(\tau=\i) \; \xlongrightarrow{\BZ_{2}^m} \; W_{\mathrm{b}}T_{\mathrm{b}}(\tau=2^{2}\i) \; \xlongrightarrow{\BZ_{2}^{e}} \; W_{\mathrm{b}}T_{\mathrm{b}}(\tau=\i),\\
    W_{\mathrm{b}}T_{\mathrm{b}}(\tau=\i) \; \xlongrightarrow{\BZ_{2}^m} \; W_{\mathrm{f}}T_{\mathrm{b}}(\tau=2^{2}\i) \; \xlongrightarrow{\BZ_{2}^{e}} \; W_{\mathrm{b}}T_{\mathrm{b}}(\tau=\i).
\end{aligned}
\end{align}
After stacking the interfaces, the action of the topological defect is written as
\begin{align}
\label{eq:cond_me_bb}
    S=\frac{\i}{2\pi}\int_{W}a_{m}\wedge \left[\d A_{L}-2(\d a_{I}-\alpha_{I}\pi w_{2})\right]+\frac{\i}{2\pi}\int_{W}a_{e}\wedge \left[2(\d a_{I}-\alpha_I\pi w_{2})-\d A_{R}\right]\,,
\end{align}
where $\alpha_{I}=0,1$ for the intermediate theory $\CT_I = W_\mathrm{b}T_\mathrm{b},W_\mathrm{f}T_\mathrm{b}$, respectively.
To specify the non-trivial action of this defect, it is sufficient to consider the junction between the magnetic gauging interface and the 't~Hooft line with charge $m$ coming from the left side.
When the fluxes of $A_{L}$ and $a_{I}$ are measured along a two-sphere in the interface around the junction, it follows from the matching condition~\eqref{eq:mag_sewing} that the constraint
\begin{align}
    m=\frac{1}{2\pi}\int_{\Sigma}F_{L}=\frac{2}{2\pi}\int_{\Sigma}(\d a_{I}-\alpha_{I}\pi w_{2})\in2\BZ
\end{align}
is required.
Thus, the magnetic charge of the 't~Hooft line must be even for consistently passing through the interface.
Otherwise, this line operator is projected out as Fig.~\ref{fig:pierce_me}.

Note that fusing some defects shown above yields the other defects in $W_{\mathrm{b}}T_{\mathrm{b}}(\tau=\i)$.
For instance, we can obtain the defect
\begin{align}
    S=\frac{2\i}{2\pi}\int_{W}a\wedge \d A_{L}-\frac{2\i}{2\pi}\int_{W}a\wedge \d a_{I}+\frac{\i}{2\pi}\int_{W}a_{I}\wedge \d A_{R}
\end{align}
by fusing the $\BS$-dual defect~\eqref{eq:s_inter_bbbb} and the condensation defect~\eqref{eq:cond_em_bb}.

\paragraph{$\bm{W_{\mathrm{b}}T_{\mathrm{f}}(\tau=\i)}$:}
We can also construct the condensation defect in $W_{\mathrm{b}}T_{\mathrm{f}}$, based on the two operations as
\begin{align}
\begin{aligned}
    W_{\mathrm{b}}T_{\mathrm{f}}(\tau=\i) \; \xlongrightarrow{\BZ_{2}^m} \; W_{\mathrm{b}}T_{\mathrm{b}}(\tau=2^{2}\i) \; \xlongrightarrow{\BZ_{2}^{e}} \; W_{\mathrm{b}}T_{\mathrm{f}}(\tau=\i),\\
    W_{\mathrm{b}}T_{\mathrm{f}}(\tau=\i) \; \xlongrightarrow{\BZ_{2}^m} \; W_{\mathrm{f}}T_{\mathrm{b}}(\tau=2^{2}\i) \; \xlongrightarrow{\BZ_{2}^{e}} \; W_{\mathrm{b}}T_{\mathrm{f}}(\tau=\i).
\end{aligned}
\end{align}
The Lagrangian of this condensation defect is 
\begin{align}
\begin{aligned}
    S=\frac{\i}{2\pi}\int_{W}a_{m}\wedge \left[\d A_{L}-2(\d a_{I}-\alpha_{I}\pi w_{2})\right]&+\frac{\i}{2\pi}\int_{W}a_{e}\wedge \left[(2\d a_{I}-\alpha_{I}\pi w_{2})-\d A_{R}\right]\\
    &-\frac{\i}{2\pi}\int_{W}(A_{L}-A_{R})\wedge \pi w_{2}.
\end{aligned}
\end{align}
This defect is a counterpart to the defect~\eqref{eq:cond_me_bb} in $W_{\mathrm{b}}T_{\mathrm{b}}(\tau=\i)$ and projects out the 't~Hooft line with an odd charge.

On the other hand, there does not exist the $\BS$-dual defect in this case, because the $\BS$-transformation maps $W_{\mathrm{b}}T_{\mathrm{f}}$ to the other theory $W_{\mathrm{f}}T_{\mathrm{b}}$.
Instead, by combining the $\BS$-transformation and the gauging procedures, we can construct a topological defect, under which $W_{\mathrm{b}}T_{\mathrm{f}}(\tau=\i)$ is self-dual.
For example, consider the following operations:
\begin{align}
\begin{aligned}
\label{eq:seq_bf_mes}
    W_{\mathrm{b}}T_{\mathrm{f}}(\tau=\i) \; \xlongrightarrow{\BZ_{2}^m} \; W_{\mathrm{b}}T_{\mathrm{b}}(\tau=2^{2}\i) \; \xlongrightarrow{\BZ_{2}^{e}} \; W_{\mathrm{f}}T_{\mathrm{b}}(\tau=\i) \; \xlongrightarrow{\BS} \; W_{\mathrm{b}}T_{\mathrm{f}}(\tau=\i),\\
    W_{\mathrm{b}}T_{\mathrm{f}}(\tau=\i) \; \xlongrightarrow{\BZ_{2}^m} \; W_{\mathrm{f}}T_{\mathrm{b}}(\tau=2^{2}\i) \; \xlongrightarrow{\BZ_{2}^{e}} \; W_{\mathrm{f}}T_{\mathrm{b}}(\tau=\i) \; \xlongrightarrow{\BS} \; W_{\mathrm{b}}T_{\mathrm{f}}(\tau=\i).
\end{aligned}
\end{align}
To specify the topological defect from these operations, we stack three appropriate interfaces.
The resulting Lagrangian of this defect is given by
\begin{align}
\begin{aligned}
    S=&\frac{\i}{2\pi}\int_{W}a_{m}\wedge\left[\d A_{L}-2(\d a_{I_{1}}-\alpha_{I_{1}}\pi w_{2})\right]-\frac{\i}{2\pi}\int_{W}A_{L}\wedge\pi w_{2}\\
    &+\frac{\i}{2\pi}\int_{W}a_{e}\wedge\left[2(\d a_{I_{1}}-\alpha_{I_{1}}\pi w_{2})-(\d a_{I_{2}}-\pi w_{2})\right]\\
    &+\frac{\i}{2\pi}\int_{W}A_{R}\wedge(\d a_{I_{2}}-\pi w_{2})+\frac{\i}{2\pi}\int_{W}A_{R}\wedge\pi w_{2}\,,
\end{aligned}
\end{align}
where $a_{m}$, $a_{e}$, $a_{I_{1}}$, and $a_{I_{2}}$ are $\mathrm{U}(1)$ gauge fields on the defect and $\alpha_{I_{1}}$ is the parameter of the second theory in the sequence of operations~\eqref{eq:seq_bf_mes}.
Since the second Stiefel-Whitney class $w_2$ on $W$ is trivial, the fluxes of $\d a_{I_{1}}-\alpha_{I_{1}}\pi w_{2}$ and $\d a_{I_{2}}-\pi w_{2}$ are not a half-integer but an integer.
This observation implies that we can redefine these fields as $(\d a_{I_{1}}-\alpha_{I_{1}}\pi w_{2})\to\d a_{I_{1}}$ and $(\d a_{I_{2}}-\pi w_{2})\to\d a_{I_{2}}$.
After integrating $a_{I_{2}}$ to set $A_{R}=a_{e}$, the Lagrangian is reduced to
\begin{align}
    S=\frac{\i}{2\pi}\int_{W}a_{m}\wedge(\d A_{L}-2\d a_{I_{1}})+\frac{2\i}{2\pi}\int_{W}A_{R}\wedge\d a_{I_{1}}-\frac{\i}{2\pi}\int_{W}(A_{L}-A_{R})\wedge\pi w_{2}\,.
\end{align}
Note that this defect is non-invertible since the 't~Hooft lines with an odd charge are projected out by the magnetic interface.
Recall that in the theory $W_{\mathrm{b}}T_{\mathrm{b}}(\tau=\i)$, there exists the invertible $\BS$-duality defect~\eqref{eq:s_inter_bbbb}. 
On the other hand, in the case of $W_{\mathrm{b}}T_{\mathrm{f}}$, we cannot construct an invertible topological defect only from the $\BS$-transformation, but one can define the non-invertible defects including the $\BS$-transformation.

\paragraph{$\bm{W_{\mathrm{f}}T_{\mathrm{b}}(\tau=\i)}$:}
In this case, the discussion of topological defects is parallel to $W_{\mathrm{b}}T_{\mathrm{f}}(\tau=\i)$.
The theory $W_{\mathrm{f}}T_{\mathrm{b}}(\tau=\i)$ has the condensation defect which corresponds to the operations
\begin{align}
\begin{aligned}
    W_{\mathrm{f}}T_{\mathrm{b}}(\tau=\i) \; \xlongrightarrow{\BZ_{2}^e} \; W_{\mathrm{b}}T_{\mathrm{b}}(\tau=\i/2^{2}) \; \xlongrightarrow{\BZ_{2}^{m}} \; W_{\mathrm{f}}T_{\mathrm{b}}(\tau=\i),\\
    W_{\mathrm{f}}T_{\mathrm{b}}(\tau=\i) \; \xlongrightarrow{\BZ_{2}^e} \; W_{\mathrm{b}}T_{\mathrm{f}}(\tau=\i/2^{2}) \; \xlongrightarrow{\BZ_{2}^{m}} \; W_{\mathrm{f}}T_{\mathrm{b}}(\tau=\i).
\end{aligned}
\end{align}
This defect has the same expression as one given in~\eqref{eq:cond_em_bb} and projects the Wilson lines with an odd charge.
Since $W_{\mathrm{f}}T_{\mathrm{b}}(\tau=\i)$ is not self-dual under the $\BS$-transformation, we combine the appropriate gauging procedures to construct the topological defects involving the $\BS$-transformation.
The resulting topological defects are non-invertible.

\subsubsection{Defects at \texorpdfstring{$\tau=\frac{1}{2}(-1+\i\sqrt{15})$}{}}
In this subsection, we study the topological defects involving the $\BT$-transformation.
In particular, we focus on the defects formed by stacking the three interfaces that are discussed in the previous subsection.
We can also construct the non-invertible symmetry of non-spin Maxwell theory at the coupling $\tau=\tfrac{1}{8}(-1+\i\sqrt{15})$ in a similar way.

\paragraph{$\bm{W_{\mathrm{b}}T_{\mathrm{b}}(\tau=\frac{1}{2}(-1+\i\sqrt{15}))}$:}
In contrast to the other two theories discussed below, this theory does not have any defect that we are focusing on.

\paragraph{$\bm{W_{\mathrm{b}}T_{\mathrm{f}}(\tau=\frac{1}{2}(-1+\i\sqrt{15}))}$:}
This theory is self-dual under the following sequence of operations:
\begin{align*}
\begin{aligned}
    W_{\mathrm{b}}T_{\mathrm{f}}(\tau=(-1+\i\sqrt{15})/2) \; &\xlongrightarrow{\BT} \; W_{\mathrm{b}}T_{\mathrm{b}}(\tau=(1+\i\sqrt{15})/2)\\
    &\xlongrightarrow{\BS} \; W_{\mathrm{b}}T_{\mathrm{b}}(\tau=(-1+\i\sqrt{15})/8) \; \xlongrightarrow{\BZ_{2}^{m}} \; W_{\mathrm{b}}T_{\mathrm{f}}(\tau=(-1+\i\sqrt{15})/2)\,.
\end{aligned}
\end{align*}
By composing appropriate interfaces, the Lagrangian of the defect is obtained as
\begin{align}
    S=\frac{2\i}{2\pi}\int_{W}A_{L}\wedge\d A_{R}+\frac{\i}{4\pi}\int_{W}A_{L}\wedge\d A_{L}-\frac{\i}{2\pi}\int_{W}(A_{L}-A_{R})\wedge\pi w_{2}\,.
\end{align}
This defect is obviously non-invertible since the magnetic gauging interface projects the 't~Hooft lines with the odd charges.

\paragraph{$\bm{W_{\mathrm{f}}T_{\mathrm{b}}(\tau=\frac{1}{2}(-1+\i\sqrt{15}))}$:}
As in the previous case, we can construct the topological defect based on the operation as
\begin{align*}
\begin{aligned}
    W_{\mathrm{f}}T_{\mathrm{b}}(\tau=(-1+\i\sqrt{15})/2) \; &\xlongrightarrow{\BT} \; W_{\mathrm{f}}T_{\mathrm{b}}(\tau=(1+\i\sqrt{15})/2)\\
    &\xlongrightarrow{\BS} \; W_{\mathrm{b}}T_{\mathrm{f}}(\tau=(-1+\i\sqrt{15})/8) \; \xlongrightarrow{\BZ_{2}^{m}} \; W_{\mathrm{f}}T_{\mathrm{b}}(\tau=(-1+\i\sqrt{15})/2)\,.
\end{aligned}
\end{align*}
The Lagrangian of this defect is given by
\begin{align}
    S=\frac{2\i}{2\pi}\int_{W}(A_{L}- \pi c)\wedge(\d A_{R}-\pi w_{2})+\frac{\i}{4\pi}\int_{W}(A_{L}-\pi c)\wedge(\d A_{L}-\pi w_{2}),
\end{align}
where $c$ is a trivialization of $w_{2}$ on $W$.
The composed topological defect turns out to be non-invertible since it includes the condensation defect that projects out half of line operators.

\section{Discussion}
\label{sec:discussion}
We have considered the Maxwell theories on a non-spin manifold. Depending on the choice of statistics for the line operators, there are three non-anomalous theories and one anomalous theory.
We established the gauging maps that connect the non-anomalous theories by coupling them to a discrete gauge theory.
We also constructed topological interfaces associated with $\mathrm{SL}(2,\BZ)$ duality and gauging of electric and magnetic one-form symmetries.
Finally, by stacking the topological interfaces, we composed various kinds of topological defects, which lead to non-invertible symmetries of non-spin Maxwell theories.

In section~\ref{sec:frac_map}, we gave the symmetry fractionalization maps in non-spin Maxwell theories.
Since the electric and magnetic one-form symmetries cannot be gauged simultaneously, we have not gauged the electric (magnetic) symmetry in the presence of the magnetic (electric) background gauge field.
For example, we do not have an electric gauging map $W_\mathrm{b}T_\mathrm{f}(\tau)\to W_\mathrm{b}T_\mathrm{b}(\tau/2^2)$ while we have its inverse by magnetic gauging
\begin{align}
\label{eq:qds_ds}
    \CF:\;W_\mathrm{b}T_\mathrm{b}(\tau) \;\longmapsto \;\frac{W_\mathrm{b}T_\mathrm{b}(\tau) \times_{\rm m} \text{BF}[0,\pi w_2]}{\BZ_2}\; \cong\; W_\mathrm{b}T_\mathrm{f}(2^2\tau)\,.
\end{align}
After gauging a symmetry, its dual symmetry typically emerges~\cite{Vafa:1989ih}, and it is conceivable that we should be able to obtain the map $W_\mathrm{b}T_\mathrm{f}(\tau)\to W_\mathrm{b}T_\mathrm{b}(\tau/2^2)$ by gauging a certain symmetry dual to the magnetic symmetry in~\eqref{eq:qds_ds}.
This general discussion implicitly assumes that an emergent symmetry does not have any 't~Hooft anomaly.
However, after the gauging~\eqref{eq:qds_ds}, the magnetic symmetry has a mixed anomaly with a background field, which obstructs returning to the original theory.
It would be interesting to consider more general setups where a quantum symmetry is anomalous after gauging.

This paper has focused on gauging a $\BZ_2$ symmetry in the electric and magnetic one-form symmetries. Its extension to $\BZ_k$ symmetry where $k$ is an odd integer is straightforward.
In this case, the $\BZ_n$ gauge theory $\text{BF}[C_1,C_2]$ coupled with Maxwell theory does not admit the background gauge fields by $C_i=\pi w_2$, and the theory after gauging always returns to the original theory $\CT$. It implies the gauging map
\begin{align}
    \CT \;\longmapsto\; \frac{\CT \times \text{BF}[0,0]}{\BZ_n} \;\cong\;\CT\,,
\end{align}
where the coupling constant is modified after gauging.
This can be done for electric and magnetic couplings due to the absence of a mixed anomaly. This map is similar to the gauging of Maxwell theory on a spin manifold and by combining the duality map in Fig.~\ref{fig:duality}, one could construct non-invertible duality defects for $\BZ_n$ gaugings.

This paper has dealt with $\mathrm{U}(1)$ gauge theory using continuous quantum field theories.
One can regularize a theory by discretizing the spacetime into a lattice, which gives a lattice gauge theory.
Our consideration was on an oriented but non-spin manifold like the complex projective plane.
We expect that the duality defects given in section~\ref{sec:top_defect} and the corresponding non-invertible symmetries appear in some lattice gauge theory after discretizing the spacetime into a lattice.
It would be helpful to extend the construction of non-invertible symmetry in pure $\BZ_2$ lattice gauge theory in~\cite{Koide:2021zxj} to a non-spin manifold.

\acknowledgments
We are grateful to K.~Ohmori, S.~Yamaguchi and R.~Yokokura for their valuable discussions.
H.\,W. acknowledges the hospitality of the particle theory and cosmology group at Tohoku University during his stay.
The work of K.\,K. was supported by FoPM, WINGS Program, the University of Tokyo and by JSPS KAKENHI Grant-in-Aid for JSPS fellows Grant No.\,23KJ0436.
The work of N.\,K. was supported by JSPS KAKENHI Grant Number JP22H01219.

\bibliographystyle{JHEP}
\bibliography{nonspin}
\end{document}